\documentclass[11pt,preprint]{aastex}
\usepackage{graphicx}
\usepackage{natbib}
\newcommand{\ha}{H$\alpha$}
\newcommand{\hb}{H$\beta$}
\newcommand{\hei}{He\,{\small I}}
\newcommand{\heii}{He\,{\small II}}
\newcommand{\ovi}{O\,{\small VI}}
\newcommand{\kms}{\,km\,s$^{-1}$}
\newcommand{\myr}{\,$M_{\sun}\,{\rm yr}^{-1}$}
\newcommand{\ro}{\,$R_{\sun}$}
\newcommand{\mo}{\,$M_{\sun}$}
\newcommand{\lo}{\,$L_{\sun}$}
\newcommand{\cmt}{\,cm$^{-3}$}

\newcommand{\ecs}{$\rm\,erg\,cm^{-2}\,s^{-1}$}
\newcommand{\ecsa}{$\rm\,erg\,cm^{-2}\,s^{-1}\,\AA^{-1}$}
\shorttitle{Transient jets in Z\,Andromedae}
\shortauthors{Skopal et al.}

\begin{document}

\title{TRANSIENT JETS IN THE SYMBIOTIC PROTOTYPE Z ANDROMEDAE}

\author{A. Skopal, 
        T. Pribulla\altaffilmark{1} 
           and 
        J. Budaj\altaffilmark{2}}
\affil{Astronomical Institute, Slovak Academy of Sciences,
       SK-059\,60 Tatransk\'{a} Lomnica, Slovakia}

\author{A. A. Vittone and L. Errico}
\affil{INAF Osservatorio Astronomico di Capodimonte,
       via Moiariello 16, I-80\,131 Napoli, Italy}

\author{M. Wolf}
\affil{Astronomical Institute, Charles University Prague,
       CZ-180\,00 Praha 8, V Hole\v sovi\v ck\'ach 2, 
       Czech Republic}

\author{M. Otsuka}
\affil{Okayama Astrophysical Observatory, NAOJ, Kamogata,
       Okayama 719-0232, Japan}

\and

\author{M. Chrastina and Z. Mikul\'{a}\v{s}ek}
\affil{Institute of Theoretical Physics and Astrophysics,
       Masaryk University Brno, Kotl\'{a}\v{r}sk\'{a} 2, 
       CZ-611\,37 Brno, Czech Republic}

\altaffiltext{1}{Visiting Astronomer: David Dunlap Observatory}
\altaffiltext{2}{2003-2006, visiting astronomer at the Department
                 of Astronomy, Pennsylvania State University}

\begin{abstract}
We present the development of the collimated bipolar jets from 
  the symbiotic prototype Z\,And that appeared and disappeared 
  during its 2006 outburst. We monitored the outburst with the 
  optical high-resolution spectroscopy and multicolor 
  $UBVR_{\rm C}$ photometry. 
In 2006 July Z\,And reached its historical maximum at $U\sim 8.0$. 
  After $\sim$1\,mag decline in mid-August, it kept its brightness 
  at a high level of $U \sim 9$ up to 2007 January. During this 
  period, rapid photometric variations with 
  $\Delta m \sim 0.06$\,mag on the timescale of hours developed. 
  Simultaneously, high-velocity satellite components appeared 
  on both sides of the \ha\ and \hb\ emission line profiles. 
  Their presence was transient, being detected to the end of 2006. 
  They were launched asymmetrically with the red/blue velocity 
  ratio of 1.2--1.3. 
  From about mid-August onward they became symmetric at 
  $\sim \pm\,1200$\kms, reducing the velocity to 
  $\sim \pm\,1100$\kms\ at their disappearance. 
  Spectral properties of these satellite emissions indicated 
  the ejection of bipolar jets collimated within an average 
  opening angle of 6$\degr.1$. 
  If the jets were expelled at the escape velocity then 
  the mass of the accreting white dwarf is 
    $M_{\rm WD} \sim 0.64$\mo. 
  We estimated the average outflow rate via jets to 
  $\dot M_{\rm jet}\sim 2\times 10^{-6}
  (R_{\rm jet}/1\,{\rm AU})^{1/2}$\myr, 
  during their August--September maximum, which 
  corresponds to the emitting mass in jets, 
  $M_{\rm jet}^{\rm em} \sim 6\times 10^{-10}
  (R_{\rm jet}/1\,{\rm AU})^{3/2}$\mo. 
  During their lifetime, the jets released the total mass of   
  $M_{\rm jet}^{\rm total} \approx 7.4\times 10^{-7}$\mo.
Evolution in the rapid photometric variability and asymmetric 
ejection of jets around the optical maximum can be explained 
by a disruption of the inner parts of the disk caused by 
radiation-induced warping of the disk. 
\end{abstract}

\keywords{stars: binaries: symbiotic -- 
          stars: individual: Z\,And -- 
          ISM: jets and outflows
         }
%
%
\section{Introduction}

Collimated jets from astrophysical objects represent very 
exciting events. They have been detected from various kinds 
of objects. A common element in all jet-producing systems is 
the central accreting star, which implies that their accretion 
disks play an important role in the formation of this type of 
mass outflow 
\citep[e.g.][ and references therein]{livio97,livio04}. 

Symbiotic stars are long-period interacting binaries (orbital 
periods are in the range of years), in which a white dwarf (WD) 
accretes material from the wind of a cool giant. This process 
generates a very hot and luminous source of radiation 
     ($T_{\rm h} \approx 10^5$\,K, 
      $L_{\rm h} \approx 10^2 - 10^4$\lo) 
and accumulates a certain amount of material onto the WD's 
surface that enlarges its effective radius to about 
0.1\ro\ during their quiescent phases. During active phases 
a large, geometrically and optically thick circumstellar disk 
with radius of a few \ro\ develops around the accretor 
\citep[][ Figs.~26 and 27]{sk05}. 
Therefore, the symbiotic stars are good candidates for producing 
high-velocity outflows in the form of collimated jets, mainly during 
outbursts. However, signatures of collimated outflows were 
indicated for only 10 from roughly 200 known symbiotics 
\citep{b+04}: seven by resolving their spatial structure 
with the radio and X-ray imaging 
\citep[e.g.][]{tay+86,crok+01,g+s04} and three by the optical 
spectroscopy \citep[e.g.][]{t+00,schmid+01}. 
Nevertheless, jets in symbiotic 
binaries are investigated very intensively from X-ray to radio 
wavelengths to understand better the physical process driving 
the collimated high-velocity mass outflow from accreting 
white dwarfs \citep[e.g.][]{sok+04,sc05,kar+07,kell+07,ss07}. 

Z\,And is considered a prototype symbiotic star. Its recent 
activity started from 2000 autumn and reached the optical maxima 
in 2000 December, 2004 September and 2006 July (Fig.~1). 
During the previous well observed large eruptions (1985, 2000), 
signatures of high-velocity outflows were detected through 
the broadening of emission lines, P-Cygni type of profiles
and the extended wings of hydrogen lines 
\citep[e.g.][]{fc+95,t+03,t+08,sok+06,sk+06,sk06}, but never 
in the form of collimated bipolar jets. 
The only signature for a nonsymmetric outflow from Z\,And 
was indicated by the 5\,GHz radio map from 2001 September, 
which showed a transient jet-like extension \citep{b+04}. 
However, no counterpart in the optical spectrum could be 
recognized \citep{b+l07}. 
Recently, during the maximum of the 2006 outburst, 
\cite{sk+pr06} discovered spectral signatures of bipolar 
jets from Z\,And. Their evidence in the optical spectra 
was confirmed by \cite{b+l07} and \cite{t+07}. 

In this contribution we investigate evolution of the jet 
features from their first appearance in 2006 July to their 
disappearance at the end of 2006, determine their physical 
parameters, and discuss the disk--jet connection during 
the outburst. 
%
%
\begin{figure*}
\centering
\begin{center}
\resizebox{14cm}{!}{\includegraphics[angle=-90]{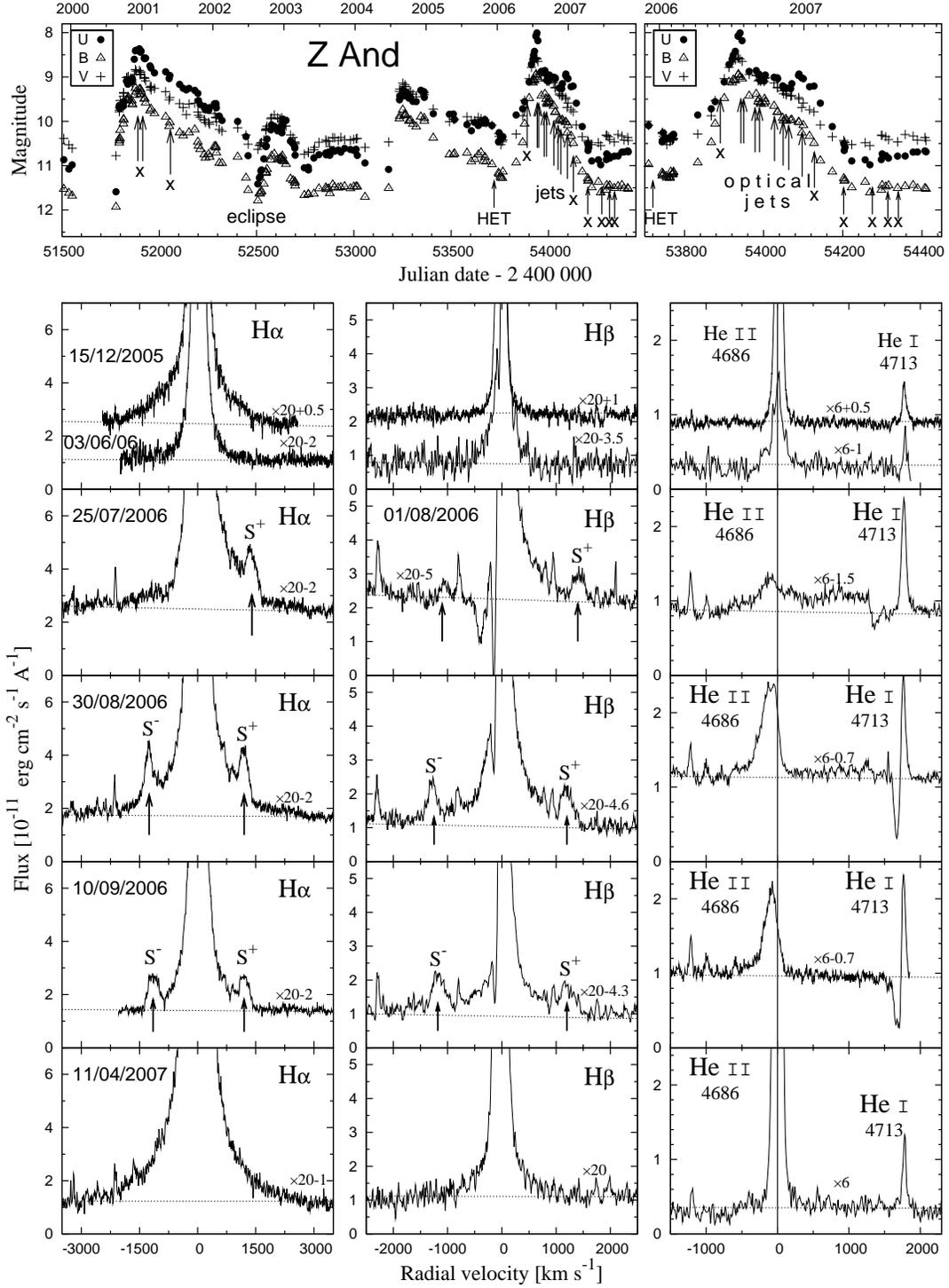}}
\caption[]{\small{
Top panels show the $U,~B$,~$V$ light curves (LCs) of Z\,And 
covering its two major eruptions that peaked in 2000 December 
and 2006 July. Arrows indicate times of spectroscopic 
observations. 
Absence of jets is denoted by "{\sf x}". 
The lower block of panels displays spectral 
regions around \ha, \hb, and \heii\,4686\,\AA\ 
during the 2006 outburst. The jet features are 
denoted by S$^{-}$ and S$^{+}$ and arrows. 
Dotted lines represent the continuum. }
          }
\end{center}
\end{figure*}
%
%
%
\begin{table}
\begin{center}
\caption{Log of spectroscopic observations}
\begin{tabular}{ccccc}
\tableline\tableline
    Date     & Julian Date & Region & $T_{\rm exp}$ &  Obs.    \\
(dd/mm/yyyy) & JD~2\,4...  &  (nm)  &   (min)       &          \\
\hline
15/12/2005 & 53720.127 & 406 - 776 & 3$\times$10  &    HET     \\
04/06/2006 & 53890.528 & 462 - 472 &     30       &   Asiago   \\
           & 53890.551 & 482 - 492 &     30       &   Asiago   \\
           & 53890.575 & 652 - 666 &     30       &   Asiago   \\
25/07/2006 & 53942.809 & 642 - 671 &    ~~5       &     DDO    \\
01/08/2006 & 53949.837 & 462 - 492 &     15       &     DDO    \\
30/08/2006 & 53978.688 & 462 - 492 &     20       &     DDO    \\
           & 53978.706 & 642 - 671 &     10       &     DDO    \\
10/09/2006 & 53989.602 & 462 - 472 &     30       &   Asiago   \\
           & 53989.626 & 482 - 492 &     30       &   Asiago   \\
           & 53989.650 & 652 - 666 &     30       &   Asiago   \\
18/10/2006 & 54027.387 & 641 - 692 &     33       & Ond\v rejov  \\
09/11/2006 & 54049.418 & 641 - 692 &     50       & Ond\v rejov  \\
23/11/2006 & 54063.428 & 641 - 692 &     50       & Ond\v rejov  \\
27/12/2006 & 54096.905 & 575 - 690 &     20       &     OAO    \\
           & 54096.919 & 575 - 690 &     20       &     OAO    \\
30/01/2007 & 54131.250 & 652 - 666 &    ~~5       &   Asiago   \\
           & 54131.263 & 652 - 666 &     30       &   Asiago   \\
           & 54131.286 & 672 - 686 &     30       &   Asiago   \\
11/04/2007 & 54202.402 & 642 - 671 &  ~~~~5.5     &     DDO    \\
22/06/2007 & 54274.697 & 642 - 671 &  ~~~~6.0     &     DDO    \\
31/07/2007 & 54313.790 & 642 - 671 &  ~~~~3.0     &     DDO    \\
27/08/2007 & 54340.485 & 652 - 686 &     30       &   Asiago   \\
\tableline
\end{tabular}
\end{center}
\end{table}

\section{Observations and data reduction}

Our observations of Z\,And during its 2006 outburst were 
carried out at different observatories. 

(i)
At the McDonald Observatory by the 9.2-m Hobby--Eberly 
telescope \citep[][ HET in Table~1]{ramsey+98} 
just prior to the rise of the major 2006 outburst. Observations 
were performed using the echelle high dispersion spectrograph 
\citep[][]{tull98} with the 316g5936 cross-disperser, 
3$\arcsec$ fiber, and 2 CCDs (4096$\times$4100) with 2$\times$1 
binning. Data were reduced using the standard {\small IRAF} 
procedures, which involved bias, bad pixels, flatfield, scattered 
light, continuum, and cosmic rays corrections. 
Wavelength calibration was done using Th-Ar lamp and has 
an error of about 0.1\kms. Observations were recorded 
in 71 orders which cover the spectral 
regions 4060--5880, 6010--7760\,\AA\ with the resolution 
of about R = 60\,000 and S/N$\sim$200 in the continuum. 
For the purpose of this paper we use only regions containing 
\ha, \hb, \heii\,4686, \hei\,4713 and the Raman-scattered 
\ovi\,6825 line. 

(ii) 
At the David Dunlap Observatory, University of Toronto (DDO) the 
high-resolution spectroscopy was performed by the single dispersion 
slit spectrograph equipped with a Jobin Yovon Horiba CCD detector 
(2048$\times$512 pixels of 13.5\,$\mu$m size; thinned back 
illuminated chip) mounted at the Cassegrain focus of the 1.88-m 
telescope. 
%
%
The resolution power was 12\,000 and 8\,000 
around the \ha\ and \heii\,4686$ - $\hb\ region, 
respectively. Basic treatment of the spectra was done using the 
{\small IRAF}-package software. During each night two very 
different exposures were applied to obtain a well defined \ha\ 
profile and the continuum. 
Telluric sharp absorptions in the \ha\ region were eliminated
with the aid of simultaneously observed standard star 50\,Boo. 

(iii) 
At the Asiago Astrophysical Observatory (Asiago). Here the 
high-dispersion spectroscopy was secured by the {\small REOSC} 
echelle spectrograph equipped with a {\small AIMO E2VCCD47-10} 
back illuminated CCD detector 
(1100$\times$1100 pixels of 13\,$\mu$m size) mounted at the 
Cassegrain focus of the 1.82-m telescope at Mt.~Ekar. 
The resolution power with the 200\,$\mu$m width slit was 
approximately 25\,000 and 17\,000 at the \ha\ and 
\heii\,4686 - \hb\ region, respectively. 
%
%
The spectroscopic data were processed with the {\small MIDAS} 
software package and a software developed at the Astronomical 
Observatory of Capodimonte in Napoli.
%

(iv)
At the Ond\v{r}ejov Observatory (Ond\v rejov) the high-resolution 
spectroscopy was performed by using the coude single-dispersion 
slit spectrograph of 2-m reflector and the {\small BROR} CCD 
camera with the SITe-005 800$\times$2030\,pixels chip. 
The resolution power at the H$\alpha$ region was 10\,000. 
Standard initial reduction of CCD spectra 
(bias subtraction, flat-fielding and wavelength calibration) 
was carried out using modified {\small MIDAS} and {\small IRAF} 
packages. Final processing of the data was done with the aid 
of the {\small SPEFO}-package software developed at 
the Ond\v{r}ejov Observatory \citep[][]{horn,skoda}. 

(v)
At the Okayama Astrophysical Observatory (OAO) the high-dispersion 
spectra were secured with the high-dispersion echelle spectrograph 
\citep[][]{i99} at the f/29 coud\'{e} focus of the 1.88-m telescope. 
The dimension of the CCD ({\small EEV} 42-80) was 2048$\times$4096 
pixels of 13.5\,${\rm \mu m}^2$. The red cross-disperser was used. 
%
%
The spectral resolving power was 
54\,000 around the \ha\ region. The reduction and analysis 
was performed with the {\small IRAF}-package software. 
Two similar exposures were co-added to obtain a better 
signal-to-noise ratio (S/N). 

(vi)
At the Skalnat\'{e} Pleso and Star\'{a} Lesn\'{a} (pavilion G2) 
Observatories, classical photoelectric $UBVR_{\rm C}$ measurements 
were carried out by single-channel photometers mounted in the 
Cassegrain foci of 0.6-m reflectors \citep[see][ in detail]{sk+04}. 
The star BD+47\,4192 (SAO\,53150, $V$ = 8.99, $B-V$ = 0.41, 
$U-B$ = 0.14, $V-R_{\rm C}$ = 0.10; C1 in Fig.~3) was used as 
the standard star for both photoelectric and CCD observations. 
Fast CCD photometry was performed at Star\'{a} Lesn\'{a} Observatory 
using the 0.5-m telescope (pavilion G1). The {\sf SBIG ST10 MXE} 
CCD camera with the chip 2184$\times$1472 pixels and the $UBV(RI)_C$ 
Johnson-Cousins filter set were mounted at the Newtonian focus. 
The size of the pixel is 6.8\,$\mu$m and the scale 0.56$\arcsec$/pixel, 
corresponding to the field of view (FOV) of a CCD frame of about 
24$\times$16 arcmin. 

(vii)
At the MonteBoo Observatory of the Masaryk University in Brno 
(Czech Republic), additional high-time-resolution optical
photometry was obtained during two nights, 2006 September 6 and
2006 September 11. 
The observational data were carried out using {\sf SBIG ST-8} 
dual chip CCD camera and Kron-Cousins $BV(RI)_C$ filter set attached 
at the Newtonian focus of the 602/2780 telescope. The CCD chip has 
1534$\times$1020 pixels with the size of 9\,$\mu$m and the scale
2.04$\arcsec$/pixel (FOV\,$\sim$17.1$\times$11.4\,arcmin). 
All CCD frames were dark-subtracted, flat-fielded and corrected 
for cosmic rays. Heliocentric correction was applied for all data 
points. CCD photometric data were processed with the {\sf \small MIDAS}
software package and a software 
{\sf \small C-MUNIPACK}\footnote{See the project homepage 
http://c-munipack.sourceforge.net/}, developed 
at the Masaryk University in Brno. Other details of the CCD 
photometric reduction were described by \cite{pv05}. 
%
%
%
\begin{figure*}[ht]
\centering
\begin{center}
\resizebox{11cm}{!}{\includegraphics[angle=-90]{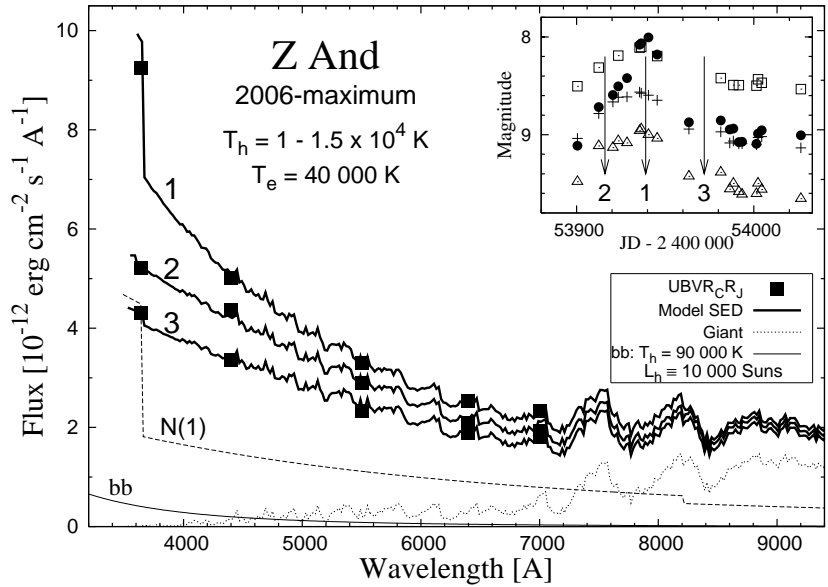}}
\caption[]{\small{
Observed (full boxes) and modeled (solid thick lines) 
SEDs at/around the 2006 maximum of Z\,And. Model for 
the red giant (dotted line) was adopted from \cite{sk05}. 
Corresponding dates (1,2,3) are shown on the $UBVR_C$ LCs 
($\bullet$, $\bigtriangleup$, +, $\Box$) inserted at the 
right-up corner. At times 2 and 3 the nebular contribution 
was negligible, while at the maximum (1) it increased by 
a factor of $\sim 20$ (denoted by N(1)). Radiation of a black 
body at the ionizing temperature of 90\,000\,K (bb) is scaled 
to a maximum of the hot object luminosity of 
$\sim 10^4$\lo\ (Sect.~3.3). }
          }
\end{center}
\end{figure*}
To verify the reality of Z\,And rapid variations during selected 
nights we measured additional comparison stars: 
C2 (GSC 03645-01592, $V$ = 11.17, $B-V$ = 0.53), 
C3 (GSC2.2~N012002338, $B\sim 12.3$\,mag) and 
C4 (SAO\,53133, $V$ = 9.17, $B-V$ = 1.36, $U-B$ = 1.11). 
The best results were achieved using the standard C1, because 
it is very similar to Z\,And during outbursts in both 
the brightness and the spectral type (see Figs.~2 and 3). 
The errors of the Z\,And -- C1 differences were mostly 
less than 0.01\,mag. 

The journal of spectroscopic observations is given in Table~1. 
A correction for heliocentric velocity was applied to all spectra. 
Arbitrary flux units were converted to absolute fluxes 
with the aid of the simultaneous $UBVR_{\rm C}$ photometry 
corrected for emission lines. The method was described by 
\cite{sk+06}. Uncertainties of such a continuum calibration 
are of a few percent for the star's brightness around 9\,mag 
in the $VR$ passbands \citep[see][ in detail]{sk07}. 
Observations were dereddened with $E_{\rm B-V}$ = 0.30 and 
resulting parameters were scaled to a distance of 1.5\,kpc 
\citep[e.g.][]{mk96}. 

\section{Analysis and results} 

\subsection{Photometric evolution around the maximum}

The top-right panel of Fig.~1 shows the $UBV$ LCs 
of Z\,And covering its recent 2006 active phase. Prior to and 
after the peak July's magnitudes, the color index $U-V\sim 0$ 
(Figs.~1 and 2). 
In July, during the maximum, the index had became temporarily 
much bluer ($U-V\sim -0.55$), which could be caused by 
a transient increase of the nebular emission. 
To understand better this behavior we modeled the $UBVR_CR_J$ 
flux points (Fig.~2)\footnote{$R_{\rm J}$ fluxes were estimated 
from the measured $R_{\rm C}$ magnitudes, corrected for the \ha\ 
emission, using transformations of \cite{b83} for the less 
red stars.} 
using the method of disentangling the composite continuum as 
described by \cite{sk05}. 
%
%
Our solutions correspond to a low temperature stellar source 
radiating at $T_{\rm h} \sim 10\,000 - 15\,000$\,K that 
dominates the optical spectrum. 
This suggests a significant contribution from a warm shell. 
At the maximum (the model 1 in Fig.~2), the SED requires 
a larger nebular contribution 
($F_{\rm stellar}/F_{\rm nebular}$ = 2.3 in $B$, emission measure 
$EM = 9.4 \times 10^{60}$\cmt\ at $T_{\rm e} = 40\,000$\,K), 
than prior to and after it ($F_{\rm stellar}/F_{\rm nebular}
\ga 23$ in $B$, $EM \sim 1.3-8.8 \times 10^{59}$\cmt). 
The range of temperatures and scaling factors in the models 
of the stellar source corresponds to its effective radius 
$R_{\rm h}^{\rm eff} = 12 \pm 4$\ro. 
We note that the absence of ultraviolet data did not allow us 
to determine the model parameters precisely. Uncertainties 
in $T_{\rm h}$ and $T_{\rm e}$ are as high as 30 and 50\%,
respectively. 

The effect of the transient increase of the nebular emission 
at the maximum could be interpreted as a result of a mass 
ejection from the active object into the particle bounded nebula, 
because new emitters will convert the excess of ionizing photons 
there into the nebular radiation. The effect of brightening
of symbiotic binary due to an increase in the mass-loss rate
was originally suggested by \cite{nv87}. 
%

In addition, we searched for a short-term photometric variability 
on the timescales of minutes to hours (Fig.~3). Just prior to 
and after the active stage we observed an irregular variation 
within $\Delta B \la 0.02$\,mag \citep[top panels of Fig.~3; 
previously measured also by][]{grom+06,s+b99}, 
whereas around/after the maximum its amplitude increased by 
a factor of $\sim$3, to $\Delta B \sim \Delta V \sim 0.06$\,mag 
(bottom panels of Fig.~3; the first detection to date). 
Their source can be identified with the aid of the corresponding 
SED. The false $(1-1.5)\times 10^4$\,K photosphere, which is 
associated with the disk, dominated the optical domain, while 
contributions from the giant and the nebula 
were negligible (Fig.~2). Therefore, the $\sim$0.06\,mag 
variations had to be produced by the disk. Also the large size 
of the disk ($R_{\rm D} > R_{\rm h}^{\rm eff} \approx 10$\ro) 
supports this interpretation -- the timescale of the rapid 
photometric fluctuations is consistent with the dynamical time 
at the outer edge of the hot stellar source. 
%
%
%
\begin{figure*}
\centering
\begin{center}
%
%
\resizebox{14.5cm}{!}{\includegraphics[angle=-90]{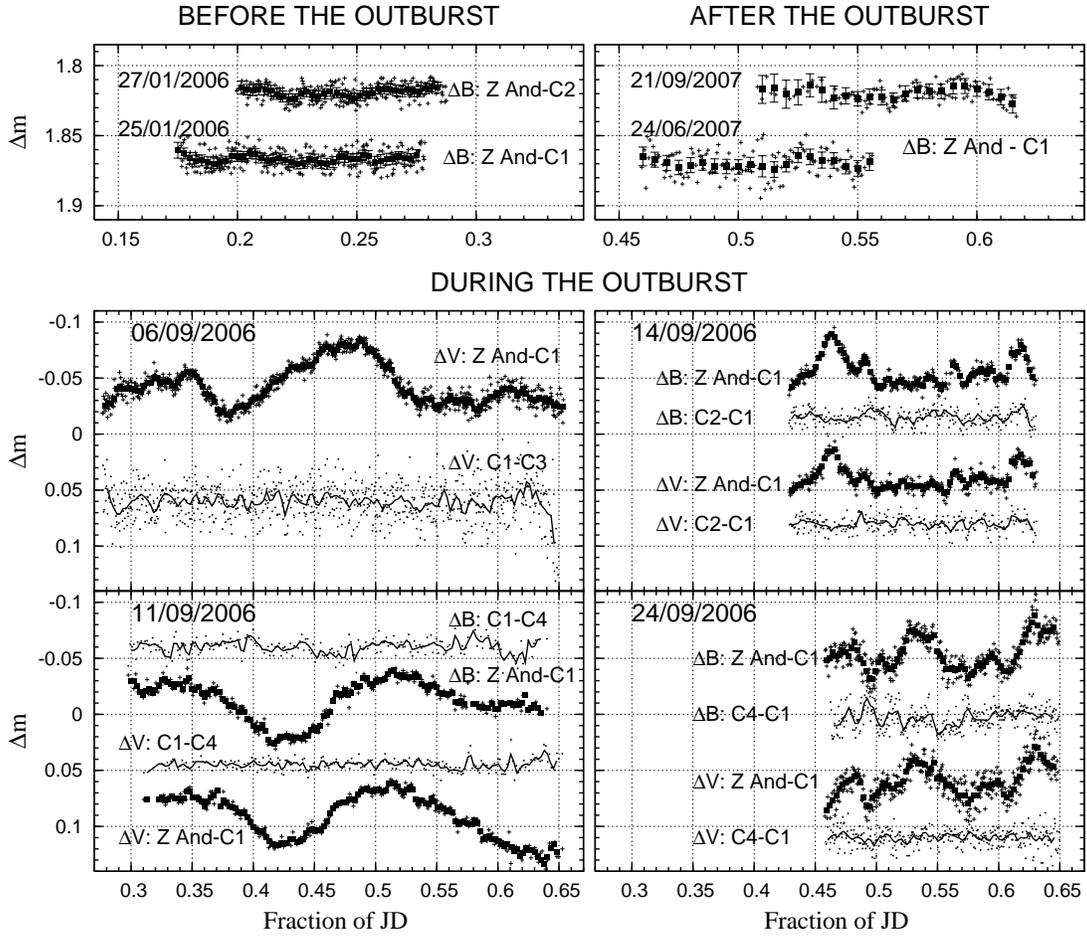}}
\caption[]{\small{
Short-term variability in the optical continuum prior to 
the 2006-outburst (top left), after it (top right) and 
around its maximum (lower panels). 
Crosses represent individual CCD measurements, while full 
squares their means within 3.6 -- 7.2 minutes. Comparison stars 
are denoted by C1, C2, C3 and C4 (see Sect.~2). 
          }}
\end{center}
\end{figure*}

\subsection{Spectroscopic evolution}

Figures~1, 4 and 5 show evolution of the \ha, \hb, \heii\,4686, 
\hei\,4713\,\AA, and the Raman-scattered \ovi\ 1032 profiles 
throughout the whole 2006-07 outburst. 

At the beginning of the 2006 active phase, prior to the 
initial rise (spectrum from 15/12/2005) as well as prior to 
the maximum (03/06/2006), the hydrogen and helium line profiles 
showed a simple emission core. At these days, emission wings of 
\ha\ extended to $\sim \pm 1\,500$ and $\sim \pm 800$\kms, 
respectively, while those of \hb\ persisted within 
$\sim \pm 500$\kms\ on both spectra. 
The emission core of the \heii\,4686 line was placed 
symmetrically with respect to its reference wavelength 
(Fig.~1). Its strong flux in the 15/12/2005 spectrum, 
F(\heii\,4686) = 2.3$\times 10^{-11}$\ecs, corresponded to 
the effective temperature of the ionizing source, 
$T_{\rm h}^{i.s.} \sim 140\,000$\,K \citep[F(\hb) = 5.5 and 
F(\hei\,4471) = 0.16$\times 10^{-11}$\ecs;][]{iijima81}.
In spite of a high temperature from nebular emission lines 
the Raman-scattered \ovi\ 1032 line disappeared entirely 
(Fig.~5). 

Our first spectrum from the maximum (25/07/2006) revealed 
a complex of strong absorptions in the \ha\ profile. One 
was extended to $\sim -1\,000$\kms\ and others cut the emission 
core at $\sim -220$\kms\ and $\sim -100$\kms\ (Fig.~4). 
In addition, a pronounced S$^{+}$ satellite component was 
present and located at +1\,385\kms. 
One week later (01/08/2006) strong absorptions at 
$\sim -400$\kms\ and $\sim -200$\kms\ appeared in 
the \hb\ and \hei\,4713\,\AA\ lines (Figs.~1 and 4). 
Two satellite components to the \hb\ emission core 
were clearly recognizable and placed asymmetrically 
at $\sim -1\,100$\kms\ and $\sim +1\,400$\kms. 
From August 30th we observed typical broad emission wings 
with the extension of about $\pm\,2\,000$ and $\pm\,1\,500$\kms\ 
in the \ha\ and \hb\ profile, respectively. They were accompanied 
by pronounced bipolarly located satellite emissions. From 
October to December their fluxes were becoming fainter and 
at the end of 2007 January only some remnants in a form of 
faint spikes could be recognized (Fig.~4). 
We attribute these satellite components to bipolar jets. 
%
%
During the presence of jets, F(\heii\,4686) fluxes decreased 
significantly (F(\heii\,4686)/F(\hb)$ \sim $0.05) and their 
profiles were shifted blueward (Fig.~1) as during the 2000-01 
maximum \citep[][]{sk+06}. The Raman line was not detectable 
\citep[see also][]{b+l07}. It became to be visible on our 
27/12/2006 spectrum as a faint redward-shifted emission 
around 6838\,\AA\ (Fig.~5). 

Our spectra taken after the maximum, from 11/04/2007 to 
27/08/2007, showed only a simple emission core of \ha\ and 
\hb\ lines with decreasing extension of their emission wings 
(Fig.~4). \heii\,4686 line became stronger, being located 
again around its reference wavelength as prior to the outburst 
(Fig.~1). The Raman emission band strengthened significantly 
and was comparable in the profile with that from quiescent 
phase (Fig.~5). 
%
%
\begin{figure*}
\centering
\begin{center}
\resizebox{14.5cm}{!}{\includegraphics[angle=-90]{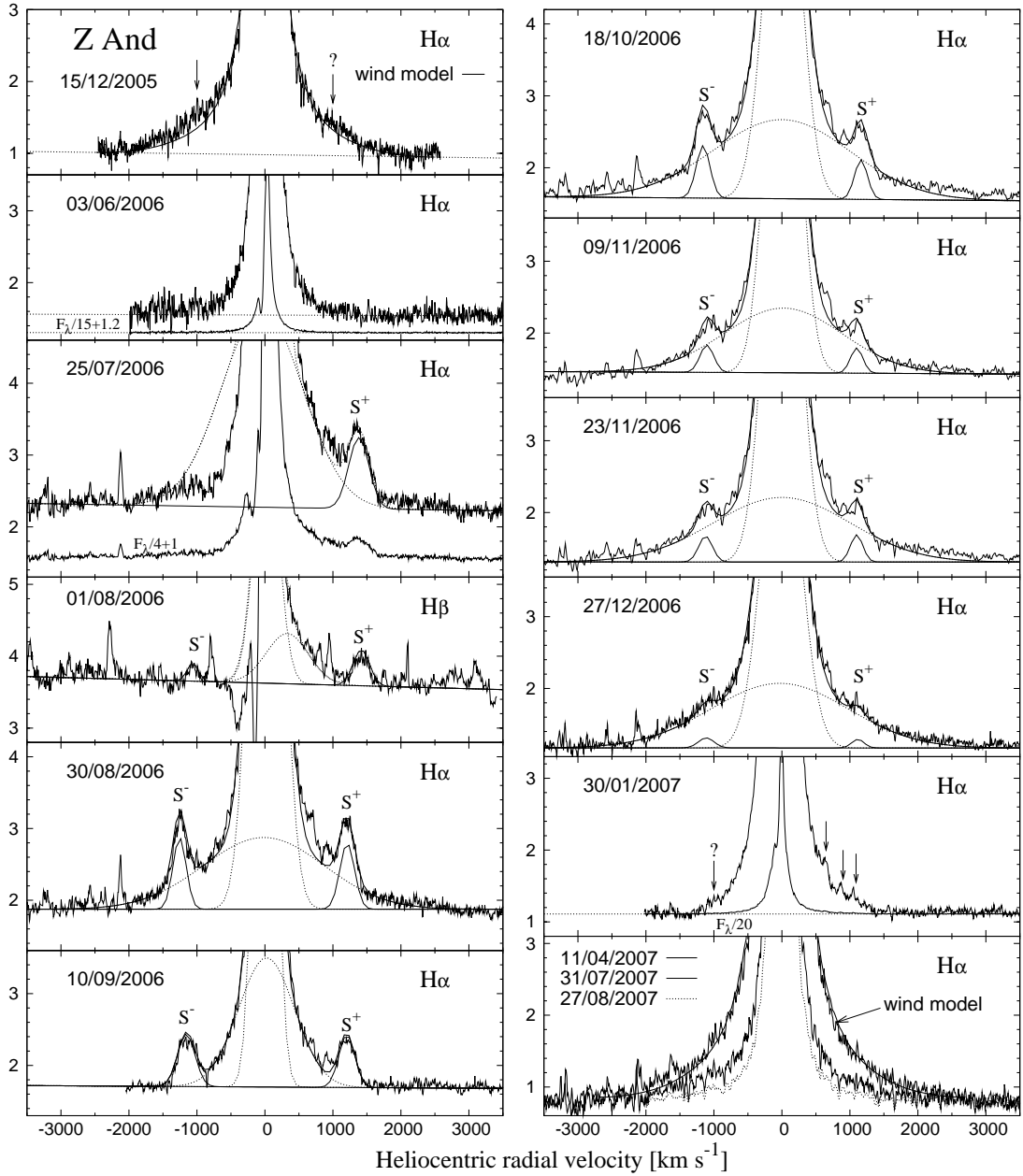}}
\caption[]{\small{
Evolution in the \ha\ (\hb) broad wings along 
the 2006 outburst. 
The jet emission components (S$^{-}$, S$^{+}$, solid thin lines) 
developed during the optical maximum and were observable to the end 
of 2006 December. We isolated them from the total profile by 
fitting its central emission with two Gaussian functions 
(dotted curves). The wind model of the broad \ha\ wings (Sect.~3.9) 
is shown here for clarity only to the spectrum from 15/12/2005 
and 11/04/2007. Corresponding parameters are in Table~2. 
Fluxes are in 10$^{-12}$\ecsa. 
          }}
\end{center}
\end{figure*}

\subsection{A disk-like structure of the hot object 
            and its evolution}

Observed SEDs from the optical maximum require a very low 
temperature of the hot object, $T_{\rm h} \sim 1-1.5\times 
10^4$\,K (Sect.~3.1). In contrast, a significantly higher 
temperature of the ionizing source, 
$T_{\rm h}^{i.s.} \sim 140\,000$ and 90\,000\,K, 
was derived from the hydrogen and helium emission 
lines prior to the outbursts (Sect~3.2) and during its 
maximum \citep[][]{b+l07}, respectively. 
Thus the hot active object was characterized by the two-temperature 
type of spectrum with $T_{\rm h} \ll T_{\rm h}^{i.s.}$ as during 
the 2000 maximum. This situation can be explained by a disk-like 
structure of the hot object viewed under a high inclination 
angle. Then the disk occults the hot ionizing source in the line 
of sight, while the nebula above/below the disk can easily be 
ionized \citep[see Sect.~4.1. of][ in detail]{sk+06}. 

Another effect resulting from the disk-structured hot object 
is the blueward shift of the \heii\,4686\,\AA\ profile 
that develops during the outbursts, while out of active phases 
it is placed symmetrically with respect to the reference 
wavelength \citep[Fig.~1 here and Fig.~2 in][]{sk+06}. 
The size of the He$^{++}$ zone was probably very small at 
the maximum (note the very low flux of the \heii\,4686 line 
during the maximum, Sect.~3.2) and thus the disk blocked 
a fraction of its redward shifted emission in the direction 
of the observer for the orbital inclination 
of about 75$\arcdeg$. 
%

The disappearance of the Raman line during active phases 
at high $T_{\rm h}^{i.s.}$ 
\citep[Sect.~3.2 and Fig.~5 here;][]{sk+06,b+l07} can 
also be a result of the disk-like structure of the hot 
object. The scattering process acts on the neutral 
atoms of hydrogen in the wind from the giant. 
So, if the ionizing source is indeed capable of producing 
the O$^{+5}$ ions (note that the ionization potential 
$\chi$(O$^{+5}) \sim 114$\,eV requires 
$T_{\rm h}^{i.s.} \sim 114\,000$\,K \citep{mn94}), 
%
%
but no Raman line is observed, implies that the O$^{+5}$ zone 
is probably too small \citep[cf. Appendix B in][]{sk+06}, 
so that the disk blocks its radiation in directions to 
the densest parts of the neutral wind at the orbital plane. 
%
%
However, the arise of a faint {\em redward}-shifted 
Raman line on our spectra from 27/12/2006 and 30/01/2007 
(Fig.~5) can be interpreted as due to the Raman scattering in 
the outer parts of the neutral giant's wind that move {\em from} 
the original \ovi\ photons. This signals a gradual dilution of 
the disk. 
Following development of a strong Raman emission with a typical 
{\em blue}-shifted shoulder in the profile (see Fig.~5) 
reflects total dilution of the optically thick material at 
the orbital plane, because the original \ovi\ photons then 
can be scattered in the densest part of the neutral wind 
from the giant along the binary axis that moves {\em against} 
them \citep[see also][for a general interpretation of the Raman 
line profile in symbiotic stars]{schmid+99}. 
The simultaneous disappearance of the disk and jets 
(cf. Figs.~4 and 5) is consistent with the necessity of 
the disk for the presence of jets \citep[e.g.][]{livio97}. 
The disk-jet connection is discussed below in Sect.~4. 

Finally, the disk-like structure of the hot {\em active} object 
in Z\,And was indicated for the first time by the model SEDs 
from the 1985 outburst, which revealed directly the presence 
of the two-temperature UV spectrum 
\citep[Sect. 5.3.4. of][]{sk05}. 
%
%
\begin{figure}
\centering
\begin{center}
\resizebox{8.5cm}{!}{\includegraphics[angle=-90]{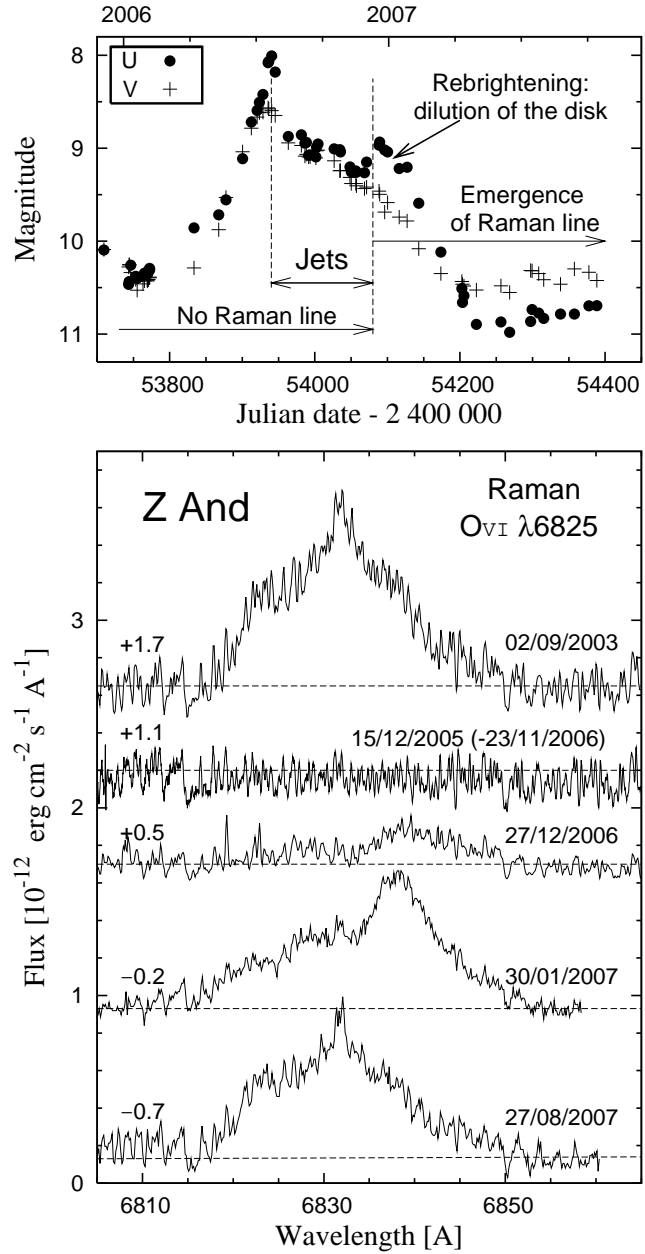}}
\caption[]{\small{
Evolution of the Raman-scattered \ovi\ 1032 line throughout 
the 2006 outburst. Top panel displays the $U$ and $V$ LCs with 
important periods in evolution of the Raman line, disk and 
jets. Bottom panel shows variation in the Raman line-profile 
along the outburst. 
Disappearance of the Raman line at high ionization temperature 
during active phases signals the presence of a large optically 
thick disk around the accretor (15/12/2005 -- 23/11/2006). 
Its emergence (from 27/12/2006) indicates the dilution of the disk 
(see Sect.~3.3 in detail). 
Numbers at the left side represent shifts with respect to 
the level of the local continuum. 
          }}
\end{center}
\end{figure}

\subsection{Measured parameters of the satellite components}

To isolate the jet components from the whole line profile, 
we formally fitted the emission line core and its extended 
foot component with two Gaussian functions. Then the residual 
jet emissions were compared with additional Gaussians. From 
their fitted parameters (the central wavelength, maximum, $I$, 
and the width $\sigma$) we derived the jet's radial velocity 
$RV_{\rm S}$, flux $F_{\rm S} = \sqrt{2 \pi}\,I\,\sigma$ 
and the width $FWHM_{\rm S} = 2\sqrt{2\ln(2)}\,\sigma$ 
(Table~2). Corresponding fits are shown in Fig.~4. 
This approach and the resolution of our spectra allowed us 
to estimate uncertainties in the $RV_{\rm S}$ values to 
10$-$23\kms\ in \ha, $FWHM_{\rm S}$ widths were possible 
to adjust to observations within 0.2$-$0.3\,\AA\ and fluxes 
within 10$-$20\% of the observed values. The smaller 
uncertainty corresponds to the stronger jet feature 
and vice versa. 

Our observations and those published by \cite{b+l07} and 
\cite{t+07} show an interesting behavior in $RV_{\rm S}$. 
From the jets launching in 2006 July to the beginning of 
2006 August, the satellite components were located 
asymmetrically, being shifted by about +150\kms\ from the 
symmetric position (the top panel of Fig.~6). 
Their average velocity was 1\,280\kms. 
Then, in the time of a few days, the jets became symmetric 
in their velocities. They had persisted at 
$|RV_{\rm S}| \sim 1\,200$\kms\ 
to about mid of September when reduced suddenly their 
$|RV_{\rm S}|$ to $\sim$1\,100\kms\ till the end of 
their detection. 
The middle panel of Fig.~6 plots the velocity ratio of the 
faster jet to its slower counterpart that reflects directly the 
evolution in the jets position. At the initial stage of the jets 
presence, this ratio was 1.2--1.3, whereas during the following 
period the velocity ratio was $\approx 1$. 
The bottom panel of Fig.~6 shows the flux ratio of the red jet 
component to the blue one, $F_{S^{+}} / F_{S^{-}}$, for 
the \ha\ line as a function of the orbital phase. 
We interpret the decrease of this ratio, indicated around 
the inferior conjunction of the giant, as an occultation 
effect (Sect.~3.8.1). 

Relatively small width of the satellite components ($FWHM_{\rm S}
\sim 250$\kms, Table~2) with respect to, for example, the broad
wings of the \ha\ line ($FWZI$(H$\alpha)\sim$ 4\,000\kms, 
e.g. Fig.~4) implies that 
   (i) the jets are highly collimated, and 
  (ii) the emitting particles can be characterized with one 
constant velocity within the visible part of the jet. 
In the following two subsections we discuss some consequences 
of these jet properties. 

\subsubsection{Collimated wind or jets?}

The winds of luminous hot stars are accelerated by 
the radiation field, produced by the underlying 
photosphere \citep[e.g.][]{lc99}. They are called line driven 
winds, because they are driven by absorption in spectral lines. 
In this process the photons transfer their energy into the kinetic 
energy of the wind, and, due to the Doppler shifts in the radial 
direction, the process acts throughout the {\em entire} wind, 
accelerating gradually its particles from the zero to the 
terminal velocity, $v_{\infty}$. As a result, we observe 
broad emission lines in the spectra of these stars. 
For symbiotic binaries, \cite{sk06} suggested that the broad 
wings of the \ha\ profiles ($v_{\infty}\ga 2\,000$\kms) are 
formed in the stellar wind from their hot components. 

The relatively narrow width of the jet 
features suggests that they are driven by a different 
mechanisms. 
If the jets were driven like the wind, their observed 
profile would be a few thousands \kms\ broad, because 
of the large velocity dispersion in the radial direction 
($0 - v_{\infty}$), and thus would hardly be detectable 
due to their low emission. 
Qualitatively, the narrow jet profiles imply that the jet engine 
works like a gun accelerating material through a {\em short} 
path (relatively to the jet length) to a high velocity, at 
which the shot-out-material continues to further distances. 
Also the connection between the disk and the jets, we discuss 
in Sect.~4, suggests that the jet launching mechanism is very 
different from that accelerating the wind. 
Generally, the jet engine transforms the kinetic energy of 
the inner parts of the disk to the kinetic energy of the flow 
in the vertical direction. In this respect, \cite{sr03} and 
\cite{sl04} suggested a scenario how jets could be blown from 
accretion disks. 
%
%

\subsubsection{Geometry of the jet emitting region}

The narrow $FWHM_{\rm S}$ allow us to assume a constant
velocity of the emitting particles within the visible
part of the jet. 
Then, if the broadening of the satellite components is caused 
by thermal motions, the corresponding Doppler width 
($\sim 250$\kms, i.e. $\sim 5.5$\,\AA\ in \ha) would require 
unrealistically high temperature of a few millions of Kelvins. 
In addition, the profiles of satellite lines have sharp edges, 
do not have typical wings, and their peaks are often 
flat (Fig.~1, e.g. 10/09/2006). 
If the jet consisted of a parallel beam of particles, i.e. 
having the geometry of a narrow column, we could not observe 
any real broadening. 
Therefore, we ascribe the observed line width to the dispersion 
in the line-of-sight velocity components of the jet particles. 
This can be satisfied if 
the jet emitting medium has the geometry of a narrow conus 
with the peak at the central object and characterized with 
a small opening angle $\theta_0$. 
%
%
%
\begin{table*}[p!ht]
\begin{center}
\scriptsize
\caption{\small{
Parameters of Gaussian fits to the jet emission components: 
Radial velocity, $RV_{\rm S}$ [km\,s$^{-1}$], flux, $F_{\rm S}$ 
[10$^{-12}$\,erg\,cm$^{-2}$\,s$^{-1}$] and $FWHM_{\rm S}$ 
[km\,s$^{-1}$]. The level of the local continuum, $F_{\rm cont}$ 
[10$^{-12}$\,erg\,cm$^{-2}$\,s$^{-1}$\,\AA$^{-1}$], opening angle 
$\theta_0$ [$\degr$], emission measure of both jets, 
$EM_{\rm jet}$ [10$^{58}$\cmt] (Eq.~5), the mass loss rate 
through jets scaled to $R_{\rm jet} = 1$\,AU, $\dot M_{\rm jet}$ 
(Eq.~(9)) and the wind, $\dot M_{\rm W}$ (Sect.~3.9) 
in $10^{-6}$\myr, are also included. 
}}
\begin{tabular}{ccccccccccccccc}
\tableline\tableline
  Date                             &
  Line                             &
\multicolumn{2}{c}{$RV_{\rm S}$}   &
\multicolumn{2}{c}{$F_{\rm S}$}    &
\multicolumn{2}{c}{$FWHM_{\rm S}$} &
$F_{\rm cont}$                     &
\multicolumn{2}{c}{$\theta_0$}     &
$EM_{\rm jet}$                     &
\multicolumn{2}{c}{$\dot M_{\rm jet}$}    &
$\dot M_{\rm W}$                   \\
dd/mm/yy                           &
                                   &
$S^{-}$                            &
$S^{+}$                            &
$S^{-}$                            &
$S^{+}$                            &
$S^{-}$                            &
$S^{+}$                            &
                                   &  
$S^{-}$                            &
$S^{+}$                            &
                                   &  
$S^{-}$                            &
$S^{+}$                            &
                                   \\
\tableline
15/12/05&\ha&\multicolumn{6}{c}{no~~jets~~detected}&1.0&-- &-- &--  & -- & -- &1.3 \\
03/06/06&\ha&\multicolumn{6}{c}{no~~jets~~detected}&1.6&-- &-- &--  & -- & -- &0.57\\
25/07/06&\ha& -1120 & +1385 & --  & 8.2 & --  & 355&2.3&-- &7.3&1.2 & -- &1.7 & -- \\
01/08/06&\hb& -1100 & +1408 & 1.1 & 1.9 & 253 & 289&3.6&6.6&5.9&1.2 &0.76&1.1 & -- \\
30/08/06&\ha& -1250 & +1214 & 5.3 & 5.5 & 227 & 255&1.9&5.2&6.0&1.6 &0.91&1.0 &2.5 \\
        &\hb& -1267 & +1212 & 2.2 & 2.3 & 204 & 225&2.8&4.6&5.3&1.8 &0.86&0.97& -- \\
10/09/06&\ha& -1152 & +1208 & 5.1 & 4.8 & 301 & 280&1.7&7.5&6.6&1.5 &1.2 &1.1 &1.5 \\
        &\hb& -1164 & +1216 & 3.1 & 2.7 & 275 & 275&2.6&6.7&6.5&2.3 &1.4 &1.3 & -- \\
18/10/06&\ha& -1156 & +1165 & 4.0 & 2.9 & 237 & 226&1.6&5.9&5.5&1.0 &0.83&0.66&2.7 \\
09/11/06&\ha& -1100 & +1100 & 2.1 & 1.7 & 237 & 215&1.5&6.1&5.6&0.56&0.59&0.49&2.4 \\
23/11/06&\ha& -1120 & +1100 & 2.0 & 1.9 & 237 & 215&1.3&6.0&5.6&0.57&0.58&0.52&2.7 \\
27/12/06&\ha& -1110 & +1110 &0.87 &0.60 & 247 & 237&1.2&6.3&6.1&0.22&0.38&0.33&2.5 \\
30/01/07&\ha&\multicolumn{6}{c}{no~~jets~~detected}&1.1 &-- &-- &--  & -- & -- &1.0 \\
11/04/07&\ha&\multicolumn{6}{c}{no~~jets~~detected}&0.83&-- &-- &--  & -- & -- &2.5 \\
22/06/07&\ha&\multicolumn{6}{c}{no~~jets~~detected}&0.80&-- &-- &--  & -- & -- &1.5 \\
31/07/07&\ha&\multicolumn{6}{c}{no~~jets~~detected}&0.84&-- &-- &--  & -- & -- &1.3 \\
27/08/07&\ha&\multicolumn{6}{c}{no~~jets~~detected}&0.82&-- &-- &--  & -- & -- &0.78 \\
\tableline
\end{tabular}
\end{center}
\normalsize
\end{table*}
%
%
%
\begin{figure}
\centering
\begin{center}
%
\resizebox{8.5cm}{!}{\includegraphics[angle=-90]{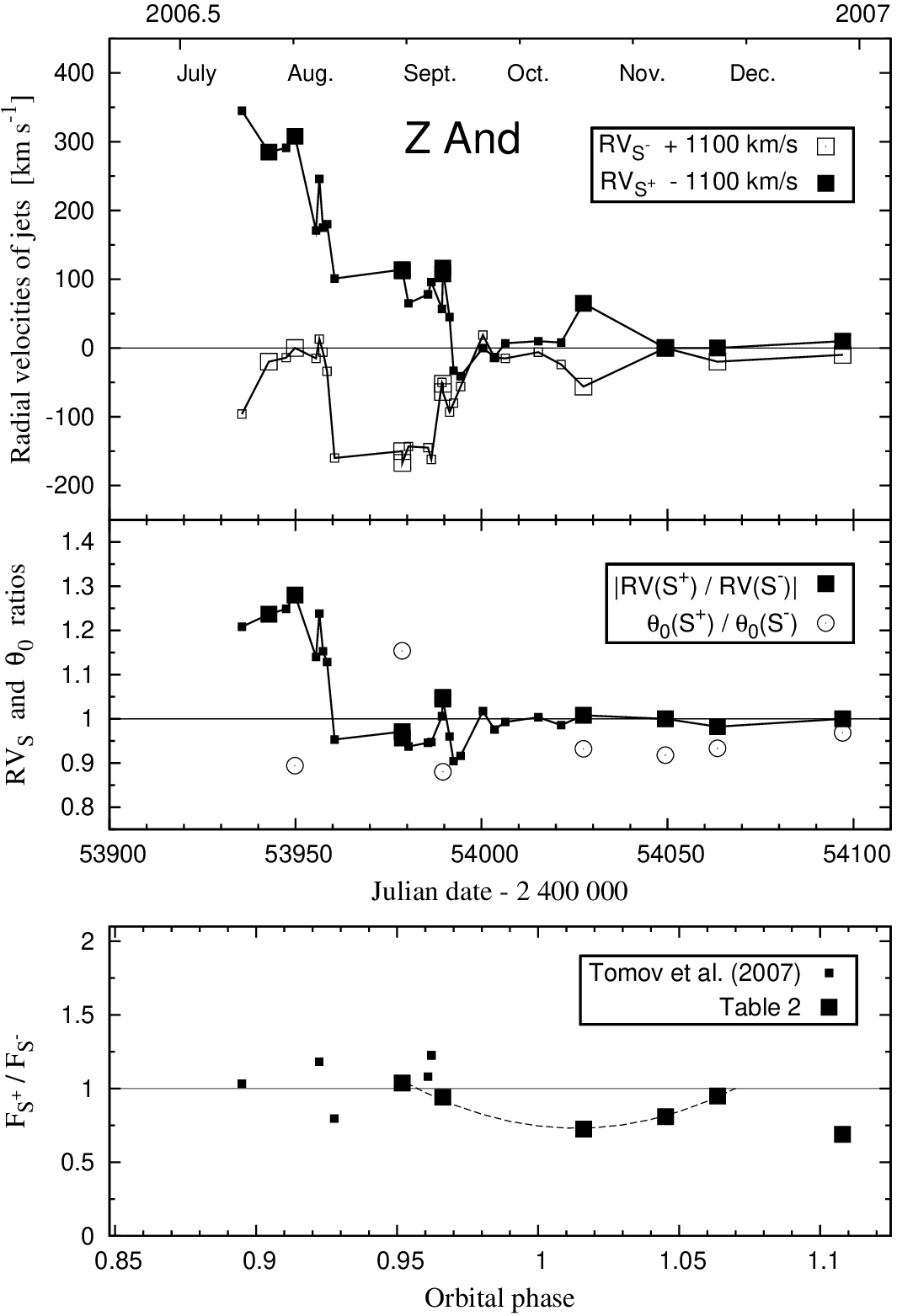}}
\caption[]{\small{
Top: 
evolution of the jet radial velocities, $RV_{\rm S}$. Large 
symbols represent our data (Table~2), while small ones are 
from \cite{b+l07} and \cite{t+07} for the \ha\ line. 
Measured values were shifted by $\pm 1\,100$\kms\ to visualize 
better their relative position. 
Middle: 
$RV_{\rm S^{+}}/RV_{\rm S^{-}}$ ratios demonstrate the 
asymmetry in the jet radial velocities at the beginning 
of their launching. Compared is the ratio of opening angles 
(Sects.~3.5. and 4.2). 
Bottom: 
the ratio of the jet component fluxes, $F_{S^{+}} / F_{S^{-}}$, 
for the \ha\ line as a function of the orbital phase. Their 
transient decrease around the phase 1.0 could be the effect 
of occultation of the red jet component by the giant 
(Sect.~3.8.1). 
          }}
\end{center}
\end{figure}

\subsection{Opening angle of the jets}

The conus geometry of the jet and the constant velocity of
its particles (Sect.~3.4.2.) allow us to express the observed
width of the satellite component as a function of its opening
angle $\theta_0$ and the orbital inclination $i$. Assuming 
that the jets were launched with the velocity, $v_{\rm jet}$, 
perpendicularly to the disk plane that coincides with the 
orbital one, the maximum dispersion of the line-of-sight jet 
velocity component corresponds to 
$2\times HWZI_{\rm S} = v_{\rm jet}\cos(i-\theta_{0}/2) - 
               v_{\rm jet}\cos(i+\theta_{0}/2) 
             = 2\sin(i)\sin(\theta_{0}/2)$, 
where the half width at the zero intensity of the jet
($HWZI_{\rm S}$) is in \kms. Then for the measured central
jet's velocity, $RV_{\rm S} = v_{\rm jet}\cos(i)$,
the opening angle can be approximated as
\begin{equation}
\theta_0 = 2\,\sin^{-1}
           \Big[\frac{HWZI_{\rm S}}{RV_{\rm S}\tan(i)}\Big].
\end{equation}
A similar relation was also used by \cite{shahbaz}. 
Corresponding parameters from Table~2, orbital inclination 
$i = 76\degr$ \citep{sk03} and adopting 
$HWZI_{\rm S} = FWHM_{\rm S}$, yield its average value as 
\begin{equation}
     \theta_0 = 6.\degr1 \pm 0.45\times \Delta i,
\end{equation}
where the uncertainty represents a total differential of 
function (1) for uncertainties in $FWHM_{\rm S}$, 
$RV_{\rm S}$ (Sect.~3.4) and that in the orbital inclination, 
$\Delta i$. 

\subsection{The jet speed and the mass of the accretor}

It is generally considered that the jet velocity in all the 
jet-producing objects is of the order of the escape velocity 
from their central stars, which indicates that the jet-type 
outflows originate from the vicinity of the accretors 
\citep[e.g.][ and references therein]{livio97}. 
Recently, \cite{sr03} and \cite{sl04} performed analytical 
estimates on how the plasma close to the central object, 
specifically in the boundary layer of the disk, could be 
accelerated to velocities larger than the local escape 
velocity. 
Our observations seem to be consistent with this scenario. 
The high velocities of jets, $v_{\rm jet} \sim 5\,000$ \kms, 
observed at large distances from the central star 
(Sec.~3.8.1), imply $v_{\rm jet}/v_{\rm escape} > 1$ 
for all possible distances of the jet ejection. 
However, from observations it is difficult to determine 
accurately the distance from the accreting body, at which 
jets are launched. 
\cite{sl04} also could not rule out the possibility that 
jets can be launched somewhat away from the accretor, up to 
$\sim 3\times$ the boundary layer radius. They noted that 
this case would require a strong perturbation in the disk. 

Therefore, for a rough estimate only, we will assume that 
the jets were launched from the WD surface at 
       $v_{\rm jet} \sim v_{\rm escape}$. 
Then the average value of 
$v_{\rm jet} = RV_{\rm S}/\cos(i) = 4\,960 \pm 90$\,\kms\ 
($RV_{\rm S} = 1\,200 \pm 22$\,\kms, Table~2, $i = 76\degr$) 
and a typical WD radius of $R_{\rm WD}$ = 0.01\ro\ correspond 
to the mass of the accretor in Z\,And, 
$M_{\rm WD} \sim 0.64\,M_{\sun}$. 
We note that this value satisfies perfectly the relation 
between mass and radius of WDs \cite[e.g.][]{st83}. 
\cite{s+s97} derived 
        $M_{\rm WD} = 0.65 \pm 0.28$\mo\ 
from the spectroscopic orbit for $i = 47\degr$. Our value of 
$M_{\rm WD}$ derived for 
$i = 76\degr$ requires just a larger total system mass 
of $\sim 3.2$\mo\ to satisfy the mass function of 0.024\mo\ 
\citep{fek+00}. 
However, in the real case, the same value of $M_{\rm WD}$ 
can be obtained for different combinations of the 
$v_{\rm jet}/v_{\rm escape}\, (> 1)$ ratio and the distance 
of the jet ejection. 

Finally, we note that the large effective radius of the active 
star during the optical maximum derived from the SED 
($R_{\rm h}^{\rm eff} \sim 12$\ro\,$\gg\, R_{\rm WD}$, 
Sect.~3.1) does not contradict to launching the jets from 
the vicinity of the WD surface, because of the disk-like 
structure of the hot active object seen under a high 
inclination angle (Sect.~3.3). 

\subsection{Radiation of jets}

In this section we derive some constraints supporting that 
the jet emission is due to the photoionization of hydrogen 
and that the medium is optically thin in the direction of 
the observer. 

(i)
According to Fig.~3 of \cite{b+l07} and the $F_{\rm S}$ fluxes
from Table~2, the average Balmer decrement
$F_{\rm S}$(H$\alpha)/F_{\rm S}$(H$\beta) \sim 2.4$, which is 
close to the theoretical value of 2.75, given by the recombination 
process at the electron temperature of 20\,000\,K 
\citep[e.g.][]{gurzadyan}. Equivalent widths presented by 
Burmeister \& Leedj\"arv were converted to fluxes with the aid 
of our continuum calibration in Table~2. 
The somewhat lower values of the observed decrement than 
the theoretical one could be caused by a partial opacity 
of the nebular jet medium in the Balmer lines. Generally,
the opacity in the \ha\ line is larger than in the \hb. 
The effect, however, should not be significant. Also the
high orbital inclination and the small value of $\theta_0$
correspond to a relatively small intersection, $l$, of 
the jet conus 
with the line of sight, and thus to a low value of the optical
depth ($\tau\,\propto {\rm opacity} \times l$), which supports
rather optically thin regime in the direction of the observer.
For a comparison, the jets produced by the symbiotic star
MWC\,560 are optically thick, because they are seen nearly
pole-on, i.e. their $l$ and thus $\tau$ are very large, which
gives rise the blueshifted jet {\em absorption} components in
the spectrum \citep[][]{schmid+01}.

(ii)
If the jet emission is due to the recombination of free electrons
with protons, then it can be created only within the ionized
fraction of the ejected material. 
In Appendix~A we calculate the extension of the jet nebula
as the distance from the central hot star, at which ionizing
photons are completely consumed by jet particles along paths 
outward from the ionizing star. According to Eq.~(A2) the jet 
radius is limited by its Str\"omgren sphere at the distance 
$r_{\rm S}$ from the ionizing source, i.e. 
\begin{equation}
 R_{\rm jet} \equiv r_{\rm S} = \left(\frac{3 L_{\rm ph}}
             {4\pi \alpha_{\rm B}(H,T_{\rm e})}\,
             \bar{n}_{\rm jet}^{-2} \right)^{1/3}, 
\end{equation}
where the parameters, $L_{\rm ph},~\alpha_{\rm B}(H,T_{\rm e}),~
\bar{n}_{\rm jet}$ are explained in Appendix~A. In Section 3.8 
and Figure 7) we demonstrate that the jets radius, 
determined independently from their observed luminosity and 
the opening angle (Eq.~(7)), agrees well with the radius of 
the Str\"omgren sphere, given by the luminosity and 
the temperature of the ionizing source. 
This result confirms that the satellite components are really 
created by recombinations within the jets, ionized by 
the central star. 

Evolution in the fluxes of jets is consistent with such 
an ionization structure. 
Between the beginning of 2006 August and the end of September,
their fluxes settled (with some fluctuations) around a maximum
\citep[][this paper]{b+l07,t+07}. No increasing trend, as could 
be expected from an increase of the emission measure of jets 
(see Eq.~(5)) due to their expansion, was observed during this 
period. This implies that the emitting mass in jets was
approximately constant, and thus suggests that the satellite
components represent only the illuminated part of jets ionized 
by the hot central object, which is bounded by the Str\"omgren 
sphere. 
Figure~9 (see Appendix~B) shows that the {\em number} of 
hydrogen ionizing photons is approximately constant for 
temperatures between about 50\,000 and 130\,000\,K 
at a fixed luminosity of a black-body source. 
Therefore, the hot star produced roughly constant flux of 
ionizing photons, and thus also relevant fluxes of jets, 
in spite that its temperature increased from 75\,000\,K to 
115\,000\,K during the above mentioned period \citep[][]{b+l07}. 
Observed fluctuations were probably caused by those in 
the luminosity of the ionizing object. 

\subsection{Mass loss through the jets}

The recombination process of the radiation by the jet plasma, 
the resulting fluxes, and the geometrical and kinematics 
parameters of the jets put some constraints to determine 
the mass loss rate through jets. 
Assuming that the jets were expelled into the solid angle 
$\Delta\Omega = 2\pi [1 - \cos(\theta_0/2)]$ of the jet 
nozzle (Sect.~3.4.2.), then the corresponding mass-loss rate, 
$\dot M_{\rm jet}$, and the mean particle concentration, 
$\bar{n}_{\rm jet}$, in the jets volume, are connected via 
the mass continuity equation as 
\begin{equation}
 \dot M_{\rm jet} = \Delta\Omega\,R^2_{\rm jet}\,\mu m_{\rm H}\,
                  \bar{n}_{\rm jet}v_{\rm jet}, 
\end{equation}
where $\mu$ is the mean molecular weight and $m_{\rm H}$ is 
the mass of the hydrogen atom. 
Total luminosity produced by jets through the recombination 
transition of the \ha\ line, $L_{\rm jet}(\rm H\alpha$), is 
related to the line emissivity, 
$\varepsilon_{\alpha} n_{\rm e}n_{\rm p}$ 
(erg\,cm$^{-3}$\,s$^{-1}$), by 
%
\begin{equation}
 L_{\rm jet}({\rm H\alpha}) = 
  \varepsilon_{\alpha}\int_{V_{\rm jet}}\! n_{\rm e}n_{\rm p}\,{\rm d}V\,
  \equiv\,\varepsilon_{\alpha}\,\bar{n}^2_{\rm jet}\,V_{\rm jet}\,=\, 
  \varepsilon_{\alpha} EM_{\rm jet},
\end{equation}
where $\varepsilon_{\alpha}$ is the volume emission 
coefficient in \ha, $n_{\rm e}$ and $n_{\rm p}$ are 
concentrations of electrons and protons and $V_{\rm jet}$ and 
$EM_{\rm jet}$ are the volume and emission measure of
the jets, respectively. 
For the optically thin medium of jets, the luminosity can be 
determined from the observed fluxes as 
$L_{\rm jet} = 4\pi d^2 \times F_{\rm S}$. 
Further, Eq.~(5) assumes a completely ionized medium 
(i.e. $n_{\rm e} = n_{\rm p} \equiv \bar{n}_{\rm jet}$), 
radiating at a constant electron temperature, i.e. 
$\varepsilon_{\alpha}$ is constant throughout the jet emitting 
volume. According to Sect.~3.4.2. we approximate its geometry 
by the conus, i.e. 
\begin{equation}
   V_{\rm jet} = \frac{1}{3}R^3_{\rm jet}\times\Delta\Omega. 
\end{equation}
Substituting Eq.~(6) into Eq.~(5), we can express the jet 
radius by means of the parameters, obtained directly from 
observations ($L_{\rm jet}$ and $\Delta\Omega$), as 
\begin{equation}
 R_{\rm jet} = \left(\frac{3 L_{\rm jet}({\rm H\alpha})}
                          {\varepsilon_{\alpha}(H,T_{\rm e}) 
                           \Delta\Omega}\, 
                           \bar{n}_{\rm jet}^{-2} \right)^{1/3}.
\end{equation}
This expression allows us to rewrite Eq.~(4) as 
\begin{equation}
  \dot M_{\rm jet} = \xi_1 \times
                     \left(\frac{\Delta\Omega}
                                {\bar{n}_{\rm jet}}\right)^{1/3}
                   \left(\frac{L_{\rm jet}}
                              {[L_{\sun}]}\right)^{2/3}
                   \frac{v_{\rm jet}}{[{\rm km\,s^{-1}}]}~
                    M_{\sun}\,{\rm yr^{-1}},
\end{equation}
and/or as a function of the radius $R_{\rm jet}$ as 
\begin{equation}
  \dot M_{\rm jet} = \xi_2 \times
                   \left(\frac{R_{\rm jet}}{[AU]}
                   \frac{\Delta\Omega}{[{\rm sr}]}
                   \frac{L_{\rm jet}}{[L_{\sun}]}\right)^{1/2}
                   \frac{v_{\rm jet}}{[{\rm km\,s^{-1}}]}~
                    M_{\sun}\,{\rm yr^{-1}}, 
\end{equation}
where the factor $\xi_1$ = $5.9\times 10^{-6}$ 
              or           $1.1\times 10^{-5}$
and $\xi_2$ = $3.6\times 10^{-9}$ 
 or         $5.9\times 10^{-9}$ 
for luminosities in \ha\ or \hb, respectively. 
Volume emission coefficients for $T_{\rm e} = 2\times 10^{4}$\,K, 
$\varepsilon_{\alpha}$ = 1.83 and $\varepsilon_{\beta} = 
0.682 \times 10^{-25}\rm erg\,cm^{3}\,s^{-1}$ \citep[][]{ost89}. 
%
%
%
\begin{figure}
\centering
\begin{center}
\resizebox{9cm}{!}{\includegraphics[angle=-90]{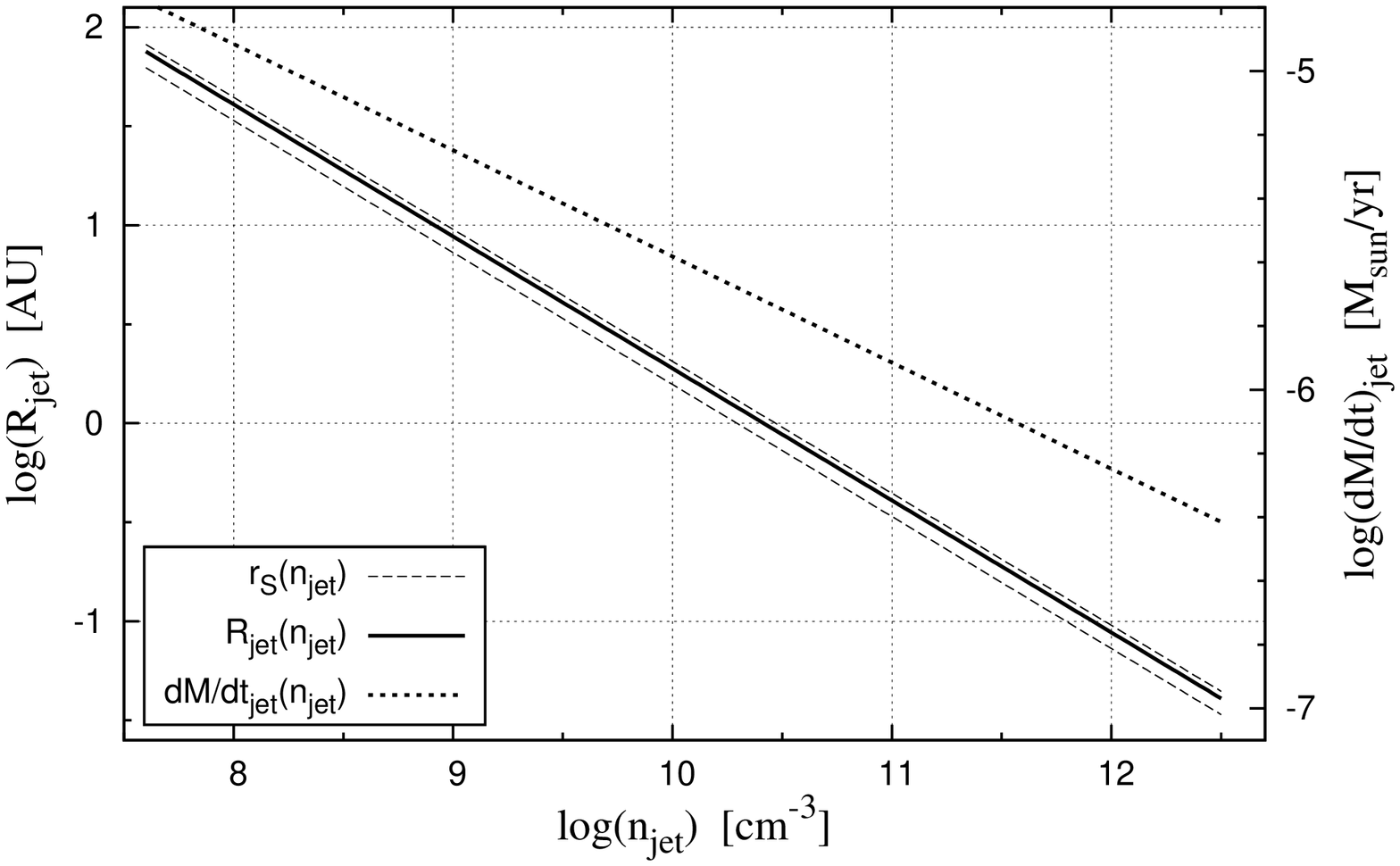}}
\caption[]{\small{
Dependencies of $R_{\rm jet}$ (Eq.~(7)) and $\dot M_{\rm jet}$ 
(Eq.~(8)) on $\bar{n}_{\rm jet}$. They were calculated for average 
values of $L_{\rm jet}$ = 0.32\lo\ (to the end of 2006 Sept.) and 
$\theta_0 = 6.\degr1$ (i.e. $\Delta\Omega = 8.9\times 10^{-3}$\,sr). 
$\dot M_{\rm jet}$ here represents the mass loss rate by both 
the jets. 
Compared are radii of the Str\"omgren sphere $r_{\rm S}$ (Eq.~(3)) 
for the hot star luminosity of $10^{4}$\lo, estimated by 
\cite{sok+06} (lower line), and for $2.2 \times 10^{4}$\lo\ 
as results for scaling this quantity to our parameters of 
$d = 1.5$\,kpc and $E_{\rm B-V}$ = 0.30 (upper line). 
A good agreement between both $R_{\rm jet}$ and $r_{\rm S}$ 
radii confirms the recombination process as responsible 
for the radiation of the jet plasma, ionized by the central 
star. 
          }}
\end{center}  
\end{figure}
%
\subsubsection{Additional constraints for $\dot M_{\rm jet}$}

According to Eqs.~(8) and (9), determination of 
$\dot M_{\rm jet}$ requires to estimate reasonable values 
of $\bar{n}_{\rm jet}$ or $R_{\rm jet}$. Figure~7 plots 
dependencies between these parameters. Below we discuss 
some constraints that allow us to determine more accurate 
ranges of the jet parameters. 

(i) 
\cite{sl04} introduced conditions for thermally launching 
jets from accretion disks around WDs. 
In their model the accreted material is strongly shocked due 
to large gradients of physical quantities in the boundary 
layer and cools on the timescale longer than its ejection 
time from the disk. The model requires large accretion 
rates of $\dot M_{\rm acc} \ga 10^{-6}$\myr. 
Some observational evidences of the disk-jet connection, 
as described in Section 4, could support this scenario. 
In agreement with this suggestion and the fact that there 
is a large disk encompassing the WD (Sects. 3.1. and 3.3.), 
we can exclude rates of 
$\dot M_{\rm jet} \lesssim 5.5\times 10^{-7}$\myr\ (i.e. 
$\dot M_{\rm acc} \lesssim 5.5\times 10^{-6}$\myr, Sect.~4.1) 
that correspond to $\bar{n}_{\rm jet} \ga 10^{12}$\cmt\ 
and $R_{\rm jet} \lesssim 0.086$\,AU (see Fig.~7), because 
of too small jet radii that would be significantly occulted 
by the disk for the orbital inclination of $\sim 75\degr$. 
Observations do not indicate this case (e.g. Table~2, Fig.~6). 
From the other side, we exclude rates 
$\dot M_{\rm jet} \ga 5.5\times 10^{-6}$\myr\ 
(i.e. $\bar{n}_{\rm jet} \lesssim 10^{9}$\cmt, 
$R_{\rm jet} \ga 9$\,AU, Fig.~7), because this would require 
too high accretion rates of $\ga 5.5\times 10^{-5}$\myr\ 
(see also Sect.~4.1). 
As the mass of the disk can be estimated to 
$\sim 10^{-4} - 10^{-5}$\mo\ (e.g. $R_{\rm D} = 20$\ro, 
$H/R_{\rm D} = 0.3$ \citep[][]{sk06} and average density 
$\log\rho$ = -7 to -8), the accretion rates of about 
$10^{-4}$\myr\ would exhaust the disk during a short 
time. However, observations indicate the presence of 
a large disk with jets for about six months, from 2006 July 
to 2007 January (Sect.~3.3, Fig.~5). 
These conditions thus limit $\dot M_{\rm jet}$ to 
$5.5\times 10^{-6} > \dot M_{\rm jet} > 5.5\times 10^{-7}$\myr\ 
and consequently, 
$10^{9} < \bar{n}_{\rm jet} < 10^{12}$\cmt\ and 
$0.1 < R_{\rm jet} < 9$\,AU. 

(ii)
During active phases of Z\,And the emission measure produced by 
the ionized hot star wind was determined to 
$EM_{\rm W} \sim 3\times 10^{59}$\cmt\ \citep[Table~1 in][]{sk06}. 
It represents the so-called low-temperature nebula, which is 
subject to eclipses in active symbiotic systems \citep[][]{sk05,sk06}, 
and thus it is located within the radius of the giant, $R_{\rm G}$, 
around the hot component. For Z\,And, the average particle 
concentration in the wind around the hot star within the radius 
$R_{\rm G} = 106$\ro\ is $\bar n_{\rm W} = 
(EM_{\rm W}/V_{\rm G})^{1/2} = 1.3 \times 10^{10}$\cmt. 
For a comparison, the mean concentration of the jet particles 
within this distance, 
$\bar n_{\rm jet}(0.5\,{\rm AU}) \sim 7.5 \times 10^{10}$\cmt, 
which corresponds to 
$\dot M_{\rm jet} \sim 1.3\times 10^{-6}$\myr\ (Fig.~7). 
This suggests that $\bar n_{\rm jet} > \bar n_{\rm W}$ 
and also the normalized mass loss rate (e.g. into 1\,sr) 
via the jets is larger than that through the wind. 

(iii)
The transient decrease in the flux ratio, 
$F_{\rm S^{+}}/F_{\rm S^{-}} < 1$, measured during the inferior 
conjunction of the giant (Fig.~6), could be caused by the occultation 
of the red jet component by the stellar disk of the giant 
\citep[also noted by][]{sw06}. 
As $2 R_{\rm G} \sim 1$\,AU, this observation suggests 
that the radii of jets are $\gtrapprox 1$\,AU. 
However, uncertainties in the orbital elements do not allow 
a more accurate estimate \citep[see][]{fl94,mk96,sk98,fek+00}. 
Finally, we note that some theoretical works, devoted to modeling 
the jets in symbiotic stars, also adopt the jet radius of 1\,AU 
\citep[e.g.][]{sc05,ss07}. 
Therefore, we determined the mass loss rate $\dot M_{\rm jet}$ 
in Table~2 for a representative value of 
$R_{\rm jet} \equiv 1$\,AU. 

\subsubsection{Emitting mass in jets}

The mean concentration and the volume of jets determine 
their emitting mass as 
\begin{equation}
   M_{\rm jet}^{em} = 
        \mu m_{\rm H}\,\bar{n}_{\rm jet}\,V_{\rm jet},
\end{equation}
which can be rewritten with the aid of Eqs.~(6) and (7) as 
\begin{equation}
 M_{\rm jet}^{em} = \xi_3 \times
           \left(\frac{\Delta\Omega}{[{\rm sr}]}
                 \frac{L_{\rm jet}}{[L_{\sun}]}\right)^{1/2}
           \left(\frac{R_{\rm jet}}{[AU]}\right)^{3/2}
                    ~M_{\sun},
\end{equation}
where $\xi_3 = 5.7\times 10^{-9}$ or 9.3$\times 10^{-9}$ for 
luminosities in \ha\ or \hb, respectively. The average value 
of $\theta_0 = 6.\degr1$ (Sect.~3.5), 
$L_{\rm jet} = 0.32\,L_{\sun}$ in \ha\ (from the 2006 
August -- September maximum) and $R_{\rm jet}$ = 1\,AU yield 
the emitting mass of 
$M_{\rm jet}^{em} = 6\times 10^{-10}~M_{\sun}$
in both jet components.
\cite{t+07} derived significantly larger value of 
$M_{\rm jet}^{em} = 1.4 \times 10^{-7}$\mo\ for the similar 
luminosity of the jet pair they observed on 12/08/2006. 
This difference results from their adopted mean density of 
only $10^{8}$\cmt, which requires very large size of one 
jet, $R_{\rm jet} \dot = 40$\,AU. However, such an extension 
of the jet is beyond the limits we estimated in Sect.~3.8. 

\subsubsection{Total mass released by jets}

With respect to our approximation, in which the jet medium 
has a constant particle density and emits within its Str\"omgren 
radius, we approximate the total mass released by the jets 
during the time of their detection, $\Delta T_{\rm jet}$, 
as 
\begin{equation}
 M_{\rm jet}^{\rm total} \approx \widehat{\dot M_{\rm jet}}
                        \times \Delta T_{\rm jet},
\end{equation}
where $\widehat{\dot M_{\rm jet}}$ is the average mass loss rate 
via jets. Values from Table~2 
($\widehat{\dot M_{\rm jet}} = 1.75\times 10^{-6}$\myr\ for the 
jet pair and $\Delta T_{\rm jet} = 154$ days) yield 
$M_{\rm jet}^{\rm total} \approx 7.4\times 10^{-7}$\mo. 

\subsection{Mass loss through the wind}

It is generally well known that during active phases the hot 
components in symbiotic binaries lose their mass in the form 
of wind. 
To estimate the rate of the mass loss produced by the active 
star through the stellar wind, we used the \ha\ method as 
suggested by \cite{sk06}. The method assumes that the broad 
wings are due to kinematics of the photoionized and optically 
thin stellar wind. 
Profiles from the maximum were modeled with the same input 
parameters as those from the 2000-maximum. The disk radius, 
$R_{\rm D} = 6.4$\ro, and the disk thickness at its edge, 
$H$ = 1.92\,\ro, were adjusted to the effective radius of 
the hot object derived from the SED assuming a flared disk 
with $H/R_{\rm D} = 0.3$. Prior to and after the outburst 
we used $R_{\rm D} = 3.5$\ro\ and $H$ = 1.05\,\ro. 
The results do not depend critically on these parameters. 
Examples of comparison between the modeled and observed 
profiles are shown in Figs.~4 and 8. Synthetic profiles fit 
well the observed wings for $|RV| \ga 300$\kms. Corresponding 
mass loss rates through the wind, $\dot M_{\rm W}$ are around 
2\,$\times\,10^{-6}\,M_{\sun}\,{\rm yr}^{-1}$ (Table~2). 
Other model parameters are the terminal velocity, which runs 
between 1\,500\kms\ and 3\,000\kms, and the acceleration 
parameter $\beta \sim 1.7$ \citep[cf. Eq.~(2) in][]{sk06}. 

It is of interest to note that the observation made just 
prior to the outbursts (the HET spectrum) revealed a more 
complex structure in the wind outflow. The $O-C$ plot shows 
a faint and very broad (FWHM $\approx$ 1\,200\kms) emission 
component superposed on the extended blue \ha\ wing (Fig.~8). 
%
%
\begin{figure}
\centering
\begin{center}
\resizebox{9cm}{!}{\includegraphics[angle=-90]{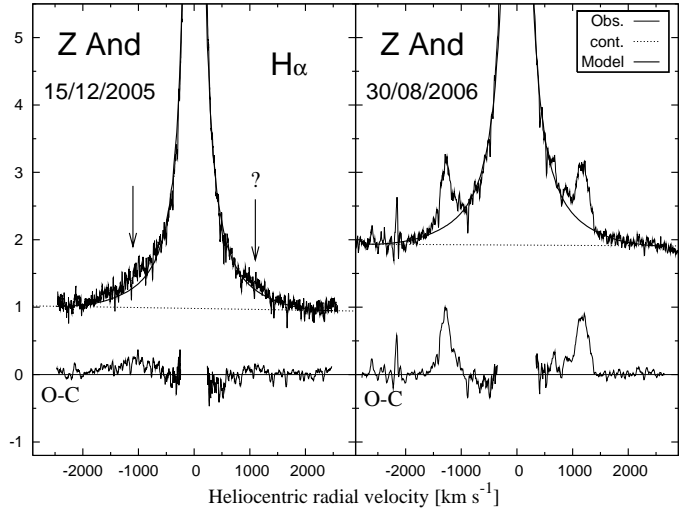}}
\caption[]{\small{
Two examples of the observed and synthetic \ha\ profiles from 
the ionized stellar wind of the hot star. The $O-C$ plot 
demonstrates a good fit for $|RV| > 300$\kms. The models 
correspond to the mass loss rate of 1.3 and 
2.5$\times\,10^{-6}$\myr\ for 15/12/2005 and 30/08/2006, 
respectively, $v_{\infty} = 2\,700$\kms\ and $\beta$ = 1.7 
(Sect.~3.9). Fluxes are in 10$^{-12}$\ecsa. 
          }}
\end{center}
\end{figure}

\section{A disk-jet connection}

\subsection{Observational evidence}

The red giant in Z\,And loses mass via the wind at 
$\dot M_{\rm G}\sim 7 \times 10^{-7}$\myr\ \citep[][ Table~3]{sk05}. 
According to the recent study on the wind accretion in binary 
stars \citep{nagae+04} the WD companion can accrete 
with efficiency up to 10\%, i.e. at $\dot M_{\rm acc}^{\rm wind} 
\sim 7 \times 10^{-8}$\myr\ during {\em quiescent phases} 
of Z\,And (the total system mass of 3.2\mo\ and the wind 
velocity of 30\kms\ at the accretor were adopted). 
This quantity corresponds to the accretion luminosity of 
$\sim 70$\lo\ (parameters from Sect.~3.6), which is at 
least a factor of 30 smaller than the observed luminosity 
of the hot stellar object \citep[e.g.][]{sk05}. This 
discrepancy led to suggestion that an additional source 
of the energy in a form of steadily burning material on the 
white dwarf surface has to be present \citep[e.g.][]{mk92}. 

During the recent active phase, the $\dot M_{\rm jet}$ rate 
put limits for the accretion rate through the disk, 
$\dot M_{\rm acc}$, because both the rates are proportional. 
Typical values of $\dot M_{\rm jet}/\dot M_{\rm acc}$ are in 
the range 0.01$-$0.3 \citep[e.g.][]{liz+88,p93,livio97}, 
although \cite{lpk03} found that $\dot M_{\rm jet}$ can almost 
be equal to $\dot M_{\rm acc}$, if the accretion energy is 
converted efficiently into magnetic energy and is emitted in 
the form of a magnetically dominated outflow or jet. 
Here we will consider the ratio 
$\dot M_{\rm jet}/\dot M_{\rm acc} \sim 0.1$. 
Accordingly, the observed 
$\dot M_{\rm jet} \sim 10^{-6}
(R_{\rm jet}/1\,{\rm AU})^{1/2}$\myr\ thus requires 
$\dot M_{\rm acc} \sim 10^{-5}$\myr. 
Two possible mechanisms how to increase the accretion rate 
are either a disk instability or an increase in the mass 
loss rate from the giant. 
If the latter possibility had been the case, the normal red 
giant in Z\,And would have increased its $\dot M_{\rm G}$ 
by 3 orders of magnitude -- from a few times 
$10^{-7}$\myr\ during quiescence to $\approx 10^{-4}$\myr\ 
for a short time during the activity, to power the observed 
jets. This seems to be unlikely. 
   Thus, assuming that $\dot M_{\rm G}$ is more or less 
constant, a disk instability had to be responsible for 
the transient, but significant, increase in $\dot M_{\rm acc}$ 
through the disk. 
We summarize critical observations supporting this suggestion 
as follows: 
\begin{enumerate}
\item
The rapid photometric variability originates in the disk 
(Sects.~3.1 and 3.3). Figure~3 shows the change from 
an irregular low-amplitude variation prior to the 
outburst to a slower, but higher-magnitude variation, 
observed during the maximum when the jets were launched. 
According to \cite{s+k03}, who examined the disk-jet 
connection in CH\,Cyg, the disappearance of the fastest 
variations indicates that the innermost disk was disrupted 
around the time when the jets were produced. This interpretation 
is based on the relationship between the disk radius and 
variability timescale \citep[slower variations come from 
larger disk radii and vice versa, as summarized by][]{s+k03}. 
Also the gas-dynamical modeling of the flow structures 
during the Z\,And outburst suggested that the sudden increase 
of the accretion rate could result from the disruption 
of the disk \citep[][]{bis+06}. 
The disk disruption, i.e. the removal of material from its 
inner parts, implies that the jets were accretion-powered. 
\item
Dramatic spectroscopic evolution around the optical maximum 
(2006 July/August), as indicated by complex absorption/emission 
profiles accompanied by the asymmetric ejection of jets 
\citep[Fig.~4, Sect.~3.2 here and Fig.~2 in][]{t+07}, could 
also be a direct result of the disk disruption. 
%
It is of interest to note that similar behavior was also 
observed during CH\,Cyg outbursts in 1982-84, 1992-94 and 
1998-00, but on somewhat longer timescales 
\citep[see Figs.~1 and 4 of][]{sk+02}. 
All the outbursts were followed by jet-extended features on 
the radio images \citep{crok+01}. However, for Z\,And these 
features have been recorded for the first time. For example, 
during its 2000-03 major outburst no jets were present 
\citep[see][]{sok+06,sk+06}. 
\item
Simultaneous presence of both the disk and jets during the 
period 2006 July -- December and their disappearance from 
2007 January (Sect.~3.3, Figs.~4 and 5) suggest that bulk of 
the accretion energy was released in the form of jets. As 
a result the hot stellar source in Z\,And had reduced its 
radiation significantly and therefore we observed a rapid 
decline in the star's brightness from 2007 January 
(see Figs.~1 and 5). 
From this point of view, the absence of jets during the previous 
(2000-03) major outburst is consistent with a slower decline 
in the brightness from the 2000-maximum (see Fig.~1) and the disk 
presence for a longer time \citep[$\sim$1.5 years,][]{sk+06}. 
In both the outbursts the presence of a large disk was 
terminated observationally by a small optical rebrightening 
accompanied by the emergence of the Raman line 
\citep[Fig.~5 here and Sect.~3.6 in][]{sk+06}. 
\end{enumerate}
We note that the disk-jet connection in Z\,And, as summarized 
in the points above, has some similarities with a hard state 
of the microquasar GRS\,1915+105 as described by \cite{belloni00} 
and later discussed by \cite{lpk03}. For the symbiotic star 
CH\,Cyg this similarity was pointed out by \cite{s+k03}. 

\subsection{Variation of jets and disk -- a self-induced 
            warping of the disk?}

The asymmetric jet episode in Z\,And lasted for less than one 
month -- from the jets launching in 2006 July to the beginning 
of 2006 August (Fig.~6, Sect.~3.4). The velocity asymmetry 
in jets was so far revealed only for other objects and at 
significantly larger timescales of years to decades 
\citep[e.g.][]{hirth+94,woitas+02,lm+03,namouni}. 
Thus, the very short duration of the asymmetric jets in Z\,And 
was probably connected with a disk instability induced during 
the optical maximum. Qualitatively, the instability could cause 
different conditions on the accretor's poles for the ejection 
of jets. Particularly, the anti-correlation of the opening angles 
to the jet velocities at the initial stage of the jets launching 
(Fig.~6) could be a possible cause of the observed jet's 
asymmetry. The larger the jet nozzle is, the lower ejection 
velocity can be expected to drive the jet flow at the same 
rate (Eq.~(4)). This possibility was already suggested by 
\cite{hirth+94} for magnetically driven jets if the opening 
angles are different due to different pressure gradients. 
%
%
The short-term, $\Delta m \sim 0.06$\,mag, photometric variations 
represent another type of variability, which developed during the 
jet ejection (Fig.~3, Sect.~3.1). Their source was associated 
with the disk (Sect.~3.1). In the previous section we ascribed 
the evolution in the short-term variability to disruption of 
the inner parts of the disk. 

The origin of both these effects could become better understood 
by investigating a disk instability due to an increase of 
the hot component luminosity at the outburst maximum. 
According to \cite{p96} irradiation of the disk by the central 
star can lead an originally planar disk to become twisted and 
tilted out of the orbital plane. In our case, additional energy 
liberated during the outburst by thermonuclear burning on the 
WD surface increases formally the efficiency $\epsilon$ 
of the accretion process \citep[see Eq.~(3.10) in][]{p97}, which 
then could lead to radiation-driven warping close to the disk's 
center even for WD accretors. For Z\,And, the radiation-induced 
warping occurs at all radii $R \ga 0.6$\ro\ for the luminosity 
of $\sim 10^4$\lo\ around the optical maximum 
\citep[e.g.][]{sok+06}, the mass accretion rate from the wind of 
$\sim 7\times 10^{-8}$\myr\ (Sect.~4.1) 
and the WD mass of 0.64\mo\ (Sect.~3.6). 
With respect to the disk radius of $\ga$10\ro\ (Sect.~3.1), 
the warping will act from inner parts of the disk. 

However, connections between jets and disk variability in 
symbiotics have not been studied theoretically in detail yet. 
Current theories on the warping and wobbling disks were 
elaborated for different types of objects and are characterized 
by significantly larger timescales than those suggested by 
observations of Z\,And \citep[e.g.][]{p96,lp97,wp99}. 

\section{Summary}

The main results of this paper can be summarized as follows: 

(i)
Between 2006 July 19 and 28 our photometric monitoring of 
the prototypical symbiotic star Z\,And revealed the highest 
maximum of its brightness that has ever been recorded by the 
multicolor photometry ($U\sim$8.0, Fig.~1). Around the mid 
of August the brightness declined to $U\sim 9$ and persisted 
around this level to 2007 January. During this period, rapid 
photometric variation ($\Delta B \sim \Delta V \sim 0.06$\,mag) 
on the timescale of hours developed (Fig.~3). 

(ii)
The SED models and spectral characteristics from the optical 
maximum can be explained by a disk-like structure of the hot 
active object (Sect.~3.3). The short-term photometric variation 
was produced by the disk (Sect.~3.1). 

(iii)
During the optical maximum a mass ejection from the active 
object was indicated photometrically through an increase 
of the nebular emission (Fig.~2) and spectroscopically 
by the development of complex absorptions on the blue side 
of the hydrogen and helium lines and a strong S$^{+}$ jet 
emission (Fig.~4). 

(iv)
High velocity satellite components to \ha\ and \hb\ emission 
lines developed in the spectrum from the optical maximum. 
Their presence in the spectrum was transient. Our first and 
last detection was on 25/07/2006 and 27/12/2006, respectively. 
Their spectral properties (Table~2) indicated ejection of highly 
collimated bipolar jets. We summarize their main characteristics 
as follows: 
\begin{enumerate}
\item
  The jets were asymmetrical in their velocities for less than 
  one month from their launching in 2006 July 
  ($RV_{\rm S^{+}}/RV_{\rm S^{-}}$ = 1.2 -- 1.3). 
  After this episode they became suddenly symmetrical until 
  the end of their presence (Fig.~6). 
\item
  The jets were collimated within an average opening angle 
  of 6$\degr$.1 for the orbital inclination 
  $i = 76\degr$ (Sect.~3.5). 
\item
  The average velocity of jets was 4\,960\kms. The ratio 
  $v_{\rm jet}/v_{\rm escape} \sim 1$ implies mass of the 
  accreting WD of $\sim 0.64\,M_{\sun}$ (Sect.~3.6). 
\item
  The average outflow rate via jets was 
$\dot M_{\rm jet}\sim 2\times 10^{-6}
(R_{\rm jet}/1\,{\rm AU})^{1/2}$\myr, 
  during their August--September maximum, which corresponds to 
  the emitting mass in jets, 
$M_{\rm jet}^{\rm em} \sim 6\times 10^{-10} 
(R_{\rm jet}/1\,{\rm AU})^{3/2}$\mo\ (Sect.~3.8.2). 
%
\item
  The total mass released by the jets during their detection 
  was approximated to 
  $M_{\rm jet}^{\rm total} \approx 7.4\times 10^{-7}\,M_{\sun}$ 
  (Sect.~3.8.3). 
\end{enumerate}

(v)
The short duration of the asymmetry in the jet velocities 
and evolution in the rapid photometric variability resulted 
from a disruption of the innermost disk. In particular, we 
pointed out a possibility of the radiation-induced warping, 
which can occur close to the disk's center due to the 
additional source of energy from the thermonuclear 
burning on the WD surface. For Z\,And parameters, 
this type of disk instability could occur at all 
radii $R \ga 0.6$\ro\ from the accretor (Sect.~4.2). 

\acknowledgments
The authors thank Mario Livio for constructive comments that 
helped improve the clarity of the paper. Dmitri Bisikalo 
and Drahom\'{\i}r Chochol are thanked for discussions. 
Martin Va\v{n}ko and Pavel Schalling are 
thanked for taking some photometric observations and Miroslav 
\v{S}lechta for initial reduction of the spectra taken at 
the Ond\v{r}ejov Observatory. 
This research was supported in part by a grant of the Slovak
Academy of Sciences No.~2/7010/27 and by the Grant Agency of 
the Czech Republic, Grant No. 205/06/0217 and 205/08/H005. 
JB acknowledges the support from the Pennsylvania State University
and the {\small NSF-NATO} grant {\small DGE-0312144} 
and the Marie Curie international reintegration grant
MIRG-CT-2007-200297. 
The Hobby-Eberly Telescope (HET) is a joint project of 
the University of Texas at Austin, the Pennsylvania State 
University, Stanford University, Ludwig-Maximillians-Universit\"at 
M\"unchen, and Georg-August-Universit\"at G\"ottingen. 
The HET is named in honor of its principal benefactors, 
William P. Hobby and Robert E. Eberly. 

\appendix
%
%
\begin{figure}[t]
\centering
\begin{center}
\resizebox{9cm}{!}{\includegraphics[angle=-90]{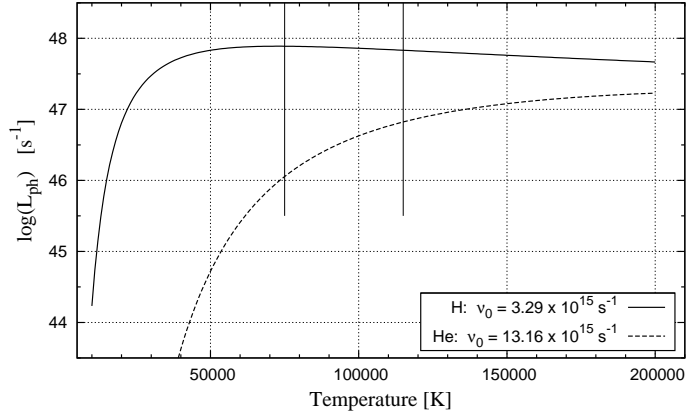}}
\caption[]{\small{
Number of photons capable to fully ionize neutral hydrogen 
and helium as a function of temperature. The photon rates 
were determined by integrating the Planck curve from the 
ionization limits $\nu_0$ to $\infty$ \citep[see Eq.~(11) 
in][]{sk01}, and scaled to the luminosity of 10\,000\,$L_{\sun}$. 
Solid thin vertical lines limit the range of temperatures 
derived from fluxes of \hb\ and \heii\,4686 during the stable 
stage of jets (Sect.~3.7). 
          }}
\end{center}  
\end{figure}

\section{Ionization boundary of jets}

Here we calculate the boundary between the ionized and neutral 
part of the jet for the case that the sources of both the 
ionizing photons and particles are located at the central star. 
The boundary is determined by the locus of points, at which 
the flux of ionizing photons is balanced by the rate of 
ionization/recombination acts inside the nebula. 
Thus the equilibrium condition between 
the number of ionizing photons $L_{\rm ph}$, emitted into 
the fraction of the sphere $\Delta\Omega /4\pi$ around 
the direction of the jet, and the number of 
recombinations, can be expressed as 
\begin{equation}
 L_{\rm ph}\frac{\Delta\Omega}{4\pi} = \frac{\Delta\Omega}{4\pi}
                \int_{0}^{R_{\rm jet}}n_{\rm e}n_{\rm p}
                \alpha_{\rm B}(H,T_{\rm e})4\pi r^2 \,{\rm d}r.
\end{equation}
It is assumed that the ionizing photons are emitted by 
the central source spherically  symmetrically, 
$\alpha_{\rm B}(H,T_{\rm e})$ stands for the total hydrogenic 
recombination coefficient in the case $B$, 
$n_{\rm e}$ and $n_{\rm p}$ are concentrations of electrons 
and protons and $r$ measures the distance from the central 
ionizing source along the jet. For the fully ionized hydrogen 
plasma ($n_{\rm e} = n_{\rm p}$) with a mean particle 
concentration, $\bar{n}_{\rm jet}$, Eq.~(A1) can be integrated 
to give 
\begin{equation}
 \frac{L_{\rm ph}}{4\pi} = \frac{1}{3}\alpha_{\rm B}(H,T_{\rm e})
                          \bar{n}_{\rm jet}^2 R_{\rm jet}^3.    
\end{equation}
In the sense of this relation, the radius of the visible part
of the jet, $R_{\rm jet}$, represents that of 
the Str\"omgren sphere. 

\section{Number of ionizing photons}

To determine the total number of photons, $L_{\rm ph}$, capable 
to ionize neutral hydrogen and helium, emitted by the star of 
temperature $T$, we integrated the Plank curve from 
the corresponding ionization limits 
(i.e. 13.598\,eV, $\nu_0 = 3.29 \times 10^{15}$\,s$^{-1}$ 
and 54.416\,eV, $\nu_0 = 13.16 \times 10^{15}$\,s$^{-1}$ for 
hydrogen and helium, respectively) to $\infty$. 
Figure~9 shows the $L_{\rm ph}$ quantity as a function of 
the temperature for a luminosity of the ionizing source 
of 10\,000\,$L_{\sun}$. It is worth to note that the function 
$L_{\rm ph}(T,H)$ is slightly decreasing from $\sim$73\,000\,K 
to higher temperatures, in spite that the total energy beyond 
the ionization limit increases. 
Therefore, irrespective to a large temperature changes 
between 50\,000 and 125\,000\,K, the number of 
the hydrogen ionizing photons varies only within 15\% of 
the maximum value at 73\,000\,K at a given luminosity 
(see Fig.~9). 

\small{

 }
\end{document}